\documentclass[twocolumn]{aastex63}

\shorttitle{On the ponderomotive force}

\usepackage{natbib}

\newcommand{\iris}{{\em IRIS}}
\newcommand{\psp}{{\em PSP}}
\newcommand{\so}{{\em SO}}

\newcommand{\identity}{\mathbf{\mathbb{I}}}

\newcommand{\longacknowledgment}{We gratefully acknowledge support by NASA grants 80NSSC18K1285, 80NSSC20K1272, 80NSSC21K0737, 80NSSC21K1684, and contract NNG09FA40C (IRIS). Resources supporting this work were provided by the NASA High-End Computing (HEC) Program through the NASA Advanced Supercomputing (NAS) Division at Ames Research Center. The simulations have been run on Pleiades through the computing project s1061, s2601, and s8305. This research is also supported by the Research Council of Norway through its Centres of Excellence scheme, project number 262622, and through grants of computing time from the Programme for Supercomputing. JMS is very grateful to his family, and especially to his wife, for their support over the many years that allowed him to progress in his career, develop Ebysus with the incredible help from the Ebysus team and culminate with this work.}

\begin{document}

\title{The Impact of Multi-fluid Effects in the Solar Chromosphere on the Ponderomotive Force Under LTE and NEQ Ionization Conditions}

\correspondingauthor{Juan Martinez-Sykora}
\email{juanms@lmsal.com}

\author[0000-0002-0333-5717]{Juan Mart\'inez-Sykora}
\affil{Lockheed Martin Solar \& Astrophysics Laboratory, 3251 Hanover Street, Palo Alto, CA 94304, USA}
\affil{Bay Area Environmental Research Institute, NASA Research Park, Moffett Field, CA 94035, USA.}
\affil{Rosseland Center for Solar Physics, University of Oslo, P.O. Box 1029 Blindern, NO-0315 Oslo, Norway}
\affil{Institute of Theoretical Astrophysics, University of Oslo, P.O. Box 1029 Blindern, NO-0315 Oslo, Norway}

\author[0000-0002-8370-952X]{Bart De Pontieu}
\affil{Lockheed Martin Solar \& Astrophysics Laboratory, 3251 Hanover Street, Palo Alto, CA 94304, USA}
\affil{Rosseland Center for Solar Physics, University of Oslo, P.O. Box 1029 Blindern, NO-0315 Oslo, Norway}
\affil{Institute of Theoretical Astrophysics, University of Oslo, P.O. Box 1029 Blindern, NO-0315 Oslo, Norway}

\author[0000-0003-0975-6659]{Viggo H. Hansteen}
\affil{Lockheed Martin Solar \& Astrophysics Laboratory, 3251 Hanover Street, Palo Alto, CA 94304, USA}
\affil{Bay Area Environmental Research Institute, NASA Research Park, Moffett Field, CA 94035, USA.}
\affil{Rosseland Center for Solar Physics, University of Oslo, P.O. Box 1029 Blindern, NO-0315 Oslo, Norway}
\affil{Institute of Theoretical Astrophysics, University of Oslo, P.O. Box 1029 Blindern, NO-0315 Oslo, Norway}

\author[0000-0002-0405-0668]{Paola Testa}
\affil{Harvard-Smithsonian Center for Astrophysics, 60 Garden Street, Cambridge, MA 02193, USA}

\author[0000-0002-8189-2922]{Q. M. Wargnier}
\affil{Lockheed Martin Solar \& Astrophysics Laboratory, 3251 Hanover Street, Palo Alto, CA 94304, USA}
\affil{Bay Area Environmental Research Institute, NASA Research Park, Moffett Field, CA 94035, USA.}

\author[0000-0002-9115-4448]{Mikolaj Szydlarski}
\affil{Rosseland Center for Solar Physics, University of Oslo, P.O. Box 1029 Blindern, NO-0315 Oslo, Norway}
\affil{Institute of Theoretical Astrophysics, University of Oslo, P.O. Box 1029 Blindern, NO-0315 Oslo, Norway}

% Mikolaj/Sam?

\begin{abstract}
The ponderomotive force is suggested to be the main mechanism to produce the so-called first ionization potential (FIP) effect — the enrichment of low FIP elements observed in the outer solar atmosphere. It is well known that the ionization of these elements occurs within the chromosphere. Therefore, this phenomenon is intimately tied to the plasma state in the chromosphere and the corona. In addition, the chromosphere is a highly complex region with a large variation in the ion-neutral collision frequencies, and hydrogen and helium ionization is largely out of equilibrium. For this study, we combine \iris\ observations, a single fluid 2.5D radiative magnetohydrodynamics (MHD) model of the solar atmosphere, including ion-neutral interaction effects and non-equilibrium (NEQ) ionization effects, and a novel multi-fluid multi-species (MFMS) numerical code (Ebysus). Nonthermal velocities of \ion{Si}{4} measured from \iris\ spectra can provide an upper limit for the strength of any high-frequency Alfvén waves. With the single-fluid model, we investigate the possible impact of NEQ ionization within the region where the FIP may occur and the plasma properties in those regions. These models suggest that regions with strong enhanced network and type II spicules are associated with the presence of large ponderomotive forces. We use the plasma properties from the single-fluid MHD model and the \iris\ observations to initialize our multi-fluid models to investigate the multi-fluid effects on the ponderomotive force associated with Alfvén waves. Our multi-fluid analysis reveals that collisions and NEQ ionization effects dramatically impact the behavior of the ponderomotive force in the chromosphere, and existing theories may need to be revisited. 
\end{abstract}


\section{Introduction} \label{sec:intro}

The chromosphere is where high FIP elements remain neutral while low FIP elements are ionized. It is here that chemical fractionation in the solar atmosphere must begin. Observations suggest that neutral elements remain well mixed with hydrogen, while ionized elements are preferentially pushed upwards \citep{Meyer:1985lr,Feldman:1992qy,Testa:2010fk,Testa:2015xi}. Other stars also exhibit fractionation effects in their coronal composition compared with photospheric abundances, and their coronal chemical composition shows some dependence on the stellar activity level (see reviews by \citealt{Testa:2010fk,Testa:2015xi}, and references therein). To further understand this process, one has to keep in mind that the chromosphere is highly dynamic because of shocks and magnetic field dynamics, and the magnitude of the magnetic field and density rapidly decreases with height \citep[e.g.,][]{Carlsson:1992kl}. In such a dynamic and highly structured chromosphere, ionization shows strong gradients (of many orders of magnitude) and is not in equilibrium. Instead, ionization depends on the history of the plasma due to the non-equilibrium effects \citep{Leenaarts:2007sf,Golding:2016wq,Martinez-Sykora:2020ApJ...889...95M}. This means that ion-neutral collision frequencies vary by many orders of magnitude on small scales within the solar atmosphere \citep[e.g.,][]{Martinez-Gomez:2015by,Martinez-Sykora:2020ApJ...889...95M,Nobreg-Siverio:2020AA...633A..66N,Khomenko:2021RSPTA.37900176K,Rempel2021ApJ...923...79R,Przybylski2022AA...664A..91P,Wargnier:2022ApJ...933..205W}. 

The chemical composition of the solar wind is also an indicator of the source region on the Sun \citep{geiss:1995SSRv...72...49G} and is critical to establishing the magnetic connectivity from the wind to the surface. A well-known example corresponds to in-situ measurements of helium abundances at 1AU that depend on the solar wind speed and phase of the activity cycle \citep{kasper:2012ApJ...745..162K}. The smallest helium abundances are observed during solar minimum. These dependencies point to mechanisms that affect the second most abundant constituent of the solar corona and whose effectiveness changes depending on the levels of activity on the Sun. Similarly, \citet{Landi2015ApJ...800..110L} have found the coronal Ne/O abundance ratio to significantly vary ($\gtrsim \times 2$, i.e., more than a factor of two) during the solar cycle and peaking at the cycle minimum. The new NASA mission Parker Solar Probe \citep[\psp,][]{Fox:2016zn} and the Solar Orbiter \citep[\so,][]{Muller2020AA...642A...1M} approach our star and measure the fields and particles much closer than 1~AU, where the plasma is less evolved and still preserves many of the original properties (composition, charge, kinetic properties, etc.). However, establishing direct links to the surface (observed with remote sensing instruments) remains a major challenge. One of the most striking recent discoveries by \psp\ is the Alfvénic magnetic field reversals known as switchbacks in the inner heliosphere \citep{Bale2021ApJ...923..174B}. The structures appear to be separated on spatial scales associated with the supergranular scales. Therefore, it is thought that these structures originate at the base of the lower solar atmosphere. A possible mechanism to drive the switchbacks is interchange reconnection in the lower corona \citep{Bale2022arXiv220807932B}. However, the switchbacks are also linked with the enrichment of alpha particles, depletion of electron temperature, and slow solar wind \citep{Fargette2021ApJ...919...96F,Woolley2021MNRAS.508..236W}. The chemical fractionation, hence the origin of the enhancement of alpha particles, should occur in or close to the chromosphere, where plasma is partially ionized, suggesting the driving mechanism of these phenomena may be connected to this region. Studies such as these highlight the challenges and opportunities of establishing the connectivity between the solar surface and the solar wind. This is important also because understanding the connectivity between stellar surfaces and their winds, the connection to the drivers, and their impact on the heliosphere and astrospheres is key to predicting space weather.

The ponderomotive force associated with strong Alfvén waves is currently regarded as the most likely mechanism to explain the FIP effect as modeled for instance by  \citet{Laming:2004qp,Laming:2015cr}. Nonetheless, this model  uses a simplified approach based on 1D semi-empirical, static atmospheres from \citet{Avrett:2007ASPC..368...81A}. These atmospheric models estimate the ionization state from averaged observations or assuming statistical equilibrium without coupling to the overlying wind. Such semi-empirical models do not capture the complex dynamics of the chromosphere and, as such, are not a good test bed for wave propagation and fractionation studies. Under those assumptions, and in order to reproduce the observables in most solar targets, the model requires Alfvén waves that are generated in the corona and that propagate (from above) into the chromosphere and that are there reflected. This theory postulates that nanoflares and reconnection in the coronal volume drive the downward propagating Alfvén waves. Observational evidence for low-frequency Alfvén waves, however, shows a predominance of upward propagating Alfvén waves that originate in the chromosphere \citep[e.g.,][]{De-Pontieu:2007bd,Okamoto:2011kx}. Similarly, observations in the corona indicate a predominance of upward propagating Alfvén waves \citep{Tomczyk:2007vn,McIntosh:2011fk}, or a mix of upward and downward waves in which the upward flux exceeds the downward flux by a significant factor \citep{McIntosh:2009lr}. It is of course possible that high frequency Alfv\'en waves, which are much more difficult to detect due to observational limitations, somehow show significantly different behavior.

\citet{Dahlburg:2016ApJ...831..160D} computed the ponderomotive acceleration from a 3D MHD model with optically thin radiation and thermal conduction. This ponderomotive acceleration in the model occurs at the footpoints of coronal loops as a byproduct of wave-driven coronal heating. The first part of this present work broadly extends this study on the ponderomotive force in realistic radiative MHD simulations, which includes non-equilibrium ionization effects and ion-neutral interaction effects (Section \ref{sec:smhd}). 

In the following, the ponderomotive force theory is briefly described in Section~\ref{sec:theo}. In  Section~\ref{sec:iris} we describe how we use IRIS observations to estimate an upper limit to the amplitude of high-frequency Alfvén waves (Section~\ref{sec:iris}). As mentioned, the results regarding the characterization of the ponderomotive force in the radiative MHD model under LTE and NEQ conditions are described in Section~\ref{sec:smhd}. We present the multi-fluid models where we investigate the collisional effects on the ponderomotive acceleration for Alfvén waves (Section~\ref{sec:mfms}). Finally, the manuscript ends with a conclusion and discussion (Section~\ref{sec:con}). 

\section{Theoretical considerations regarding the ponderomotive force} \label{sec:theo}

The Lorentz force is a key force for understanding the dynamics of plasmas. It is a straightforward term that acts on charged particles. However, it is not always practical to use this term in complex scenarios such as when electromagnetic waves or complex environments are involved. For such conditions, the ponderomotive force is often used to simplify understanding of the plasma dynamics. The ponderomotive force is based on the time average of nonlinear forces that act in the presence of oscillating electromagnetic fields. The calculation of this force is however rather cumbersome as it includes non-linear aspects, and it usually involves quite limiting assumptions, rendering the results approximate. There are a variety of formulations of the ponderomotive force, which differ in the underlying assumptions. All variations of the ponderomotive force are derived by a nonlinear dependence on the amplitude of the electric field oscillations. This force can be treated as the acceleration in a electron-ion plasma, or in ions, as nicely described in \citet{Whitelam2002SoPh..211..199W}. They described the acceleration along the magnetic field associated with the ponderomotive force resulting from nonlinear propagation of circularly polarized electro-magnetic (CPEM) waves in an electron-ion plasma as follows:  

\begin{equation}
    F = - \left(\frac{\partial}{\partial z} + \frac{k \Omega_{cj}}{\omega(\omega + \Omega_{cj})}\frac{\partial}{\partial t}\right)\frac{q_j^2|\vec{E}|^2}{m\omega(\omega+\Omega_{cj})}  ~\label{eq:pond1}
\end{equation}

\noindent where $\vec{E}$ is the electric field; $\omega$, and $k$ are the wave frequency and number, respectively; and $\Omega_{cj}$, $m_j$ and $q_j$ are the gyrofrequency, mass, and charge for species $j$ ($j$ can be any ionized species or electrons). \citet{Whitelam2002SoPh..211..199W} also considered a case in which the ion acceleration is produced by electro-magnetic electron-cyclotron (EMEC) or ion-cyclotron Alfvén (EMICA) waves in which they assumed that the ponderomotive force dominates over the background Lorentz force and pressure gradients: 

\begin{equation}
    F = - \frac{q_i^2}{m_i m_e \omega(\omega-|\Omega_{ce}|)}\left(\frac{\partial}{\partial z} - \frac{k \Omega_{ce}}{\omega(\omega - |\Omega_{ce}|)}\frac{\partial}{\partial t}\right)|\vec{E}|^2  ~\label{eq:pond1b}
\end{equation}

For the case of a multi-ions species and the ponderomotive force from EMICA waves,  the $\alpha$ species is governed by: 

\begin{eqnarray}
    F &=& - \frac{q_{i\alpha}^2}{m_{i\alpha}^2 \Omega_{ci\alpha}} \left[\frac{1}{\omega-\Omega_{ci\alpha}}\frac{\partial}{\partial z} + \right. \\ \nonumber
    &&\left. \frac{k}{\omega^2}\left(1- \frac{\Omega_{ci\alpha}^2}{(\omega-\Omega_{ci\alpha})^2}\right)\frac{\partial}{\partial t}\right]|\vec{E}|^2  ~\label{eq:pond1c}
\end{eqnarray}

Note that the spatial and temporal components are considered in the literature are sometimes considered as different forces. A nice summary of the various calculations of different forces can be found in \citet{Lundin2006SSRv..127....1L}. In short, and to list a few from that paper:  

\begin{itemize} 
\item Lundin-Hultqvist force or magnetic moment pumping (MMP) is given by: 
\begin{equation}
F = - \frac{mc^2}{2}\frac{E^2}{B^2}\nabla_{||} \ln B
\end{equation}

\citep{Lundin1989JGR....94.6665L}. Here $\vec{B}$ is the magnetic field. This ponderomotive force assumes a low frequency regime, that the electric field oscillation is perpendicular to the magnetic field and a very slow variability in $B$. It is always in the direction of decreasing B (independent of the wave propagation direction).
\item Miller force or gradient ponderomotive force 
\begin{equation}
F =  \frac{mc^2}{4 B^2}\frac{\partial E^2}{\partial z}
\end{equation}
\citep{Miller1958}. This force is related to the spatial inhomogeneity of the wave field. The derivation of the force assumes a static magnetic field with a perpendicular oscillation between the electric field and magnetic field, where the magnetic field perturbations are negligible compared to the ambient magnetic field. The formula provided is for Alfv\'en waves with low frequency. This force is positive (in direction of wave propagation) if the gradient of the wave field is positive (i.e., increasing in the direction of wave propagation). %This force assumes two extreme cases, for frequencies larger than the gyrofrequency this force is negative, i.e, opposite to the direction of the wave propagation, for a gradient that decreases in the direction of the wave. In the low frequency regime the force is oppositely directed to that of high frequency regime, i.e., positive (in the direction of wave propagation) for a gradient that decreases in the direction of the wave propagation. 
\item Abraham force is 
\begin{equation}
F = \pm \frac{mc^2}{2 v_A B^2}\frac{\partial E^2}{\partial t}. 
\end{equation}
This description of the ponderomotive force neglects standing waves and applies to waves traveling in the direction of the magnetic field. The positive (negative) sign corresponds to a wave propagating in the direction (opposite) to the magnetic field. 
\end{itemize} 

 The chromosphere is partially ionized and none of those studies considered the ponderomotive effects of damping of Alfv\'en waves from ion-neutral interactions. \citet{Haerendel1992Natur.360..241H} takes into account the presence of collisions and the ponderomotive force has a very different dependence: 

\begin{equation}
    F = \rho_n u_x^2 \frac{\omega^2}{2 v_A \nu_{ni}}  ~\label{eq:pond2}
\end{equation}

\indent where $\rho_n$, and $\nu_{ni}$ are the neutral density, and collision frequency between ions and neutrals, respectively. $v_A$ is the Alfvén speed. 
Note that this force is a factor of two smaller than originally reported by \citet{Haerendel1992Natur.360..241H}. The corrected formula is provided in \citet{De-Pontieu1998lr}. This formula is based on the WKB approximation of a weakly damped wave and is averaged over a wave period. A derivation for strongly damped waves is provided by \citep{Song:2011ly}. The nature of this force is such that it is proportional to the damping of the wave field along the direction of propagation, i.e., $F \sim - \frac{\partial E^2} {\partial z}$. Note that the frequency dependence is different from Eq.~\ref{eq:pond1}, which decreases with increasing frequency. The ion-neutral damping of the waves leads to a damping length $L$ for the wave field:

\begin{equation}
L = \frac{2 v_A \nu_{ni} \rho_t}{\omega^2 \rho_n} ~\label{eq:damp}
\end{equation}

 \indent where $\rho_t$ is the total mass density, and $\rho_n$ is the neutral mass density.

%\begin{equation}
%F \sim  - \rho_n \frac{\omega^2}{2 v_A \nu_{ni}}\frac{\partial E^2}{\partial z}    
%\end{equation}

%{\jms I hope I didn't make any mistake in the last =, please check Bart, it is not written in that way neither in Haerendel's and yours papers.}
% Note that the dependence in frequency is opposite to Eq.~\ref{eq:pond1}. This equation has considered XXX assumptions. \cite{de-Pontieu:1998lr}

As we will show with our multi-fluids models (Section~\ref{sec:ebres}), the many assumptions underlying some of these equations are not necessarily easy to justify for solar conditions or even the simplified scenarios we consider here (no density stratification, dynamics, and only Alfvén waves in the low-frequency regime).

\section{IRIS Observations: non-thermal velocity}~\label{sec:iris}

We have selected a decaying AR with plage and a large presence of spicules near the limb to characterize the high-frequency Alfvén waves traveling along the spicules, which will be used to constrain our models in the following sections. The exposure time of the selected \iris\ observation is such as can add a constrain to the high-frequencies, i.e., as short as possible (see below for details). 

\subsection{Target and data calibration}
For this study, we use \iris\ observations acquired from 2014-12-11 22:39:14 to 23:32:274 in a plage region near the limb  $(x,y)=(-910, -233)$\arcsec. \iris\ observed this active region (AR) with a very large sparse 64-step raster, with no roll, exposures of 2~s, a cadence of 3.2~s, raster-scans of 199~s, and 2x2 binning (i.e., a spatial plate scale of 0.33\arcsec and spectral sampling of 5.4~km~s$^{-1}$), using OBSID 3820105390. The observation includes fourteen full rasters. This study focuses only on the 13$^{\text{th}}$ raster, which is free of radiation hits from passage through the South Atlantic Anomaly (SAA). The slit jaw image (SJI) data are obtained in the 1330 and 2796 passbands. We use \iris\ calibrated level 2 data, which have been processed for dark current, flat field, and geometrical corrections \citep{De-Pontieu:2014fv}. 

\subsection{Analysis of the IRIS observations}~\label{sec:irisanalysis}
\begin{figure*}
    \label{fig:obsntw}
\includegraphics[width=0.97\textwidth]{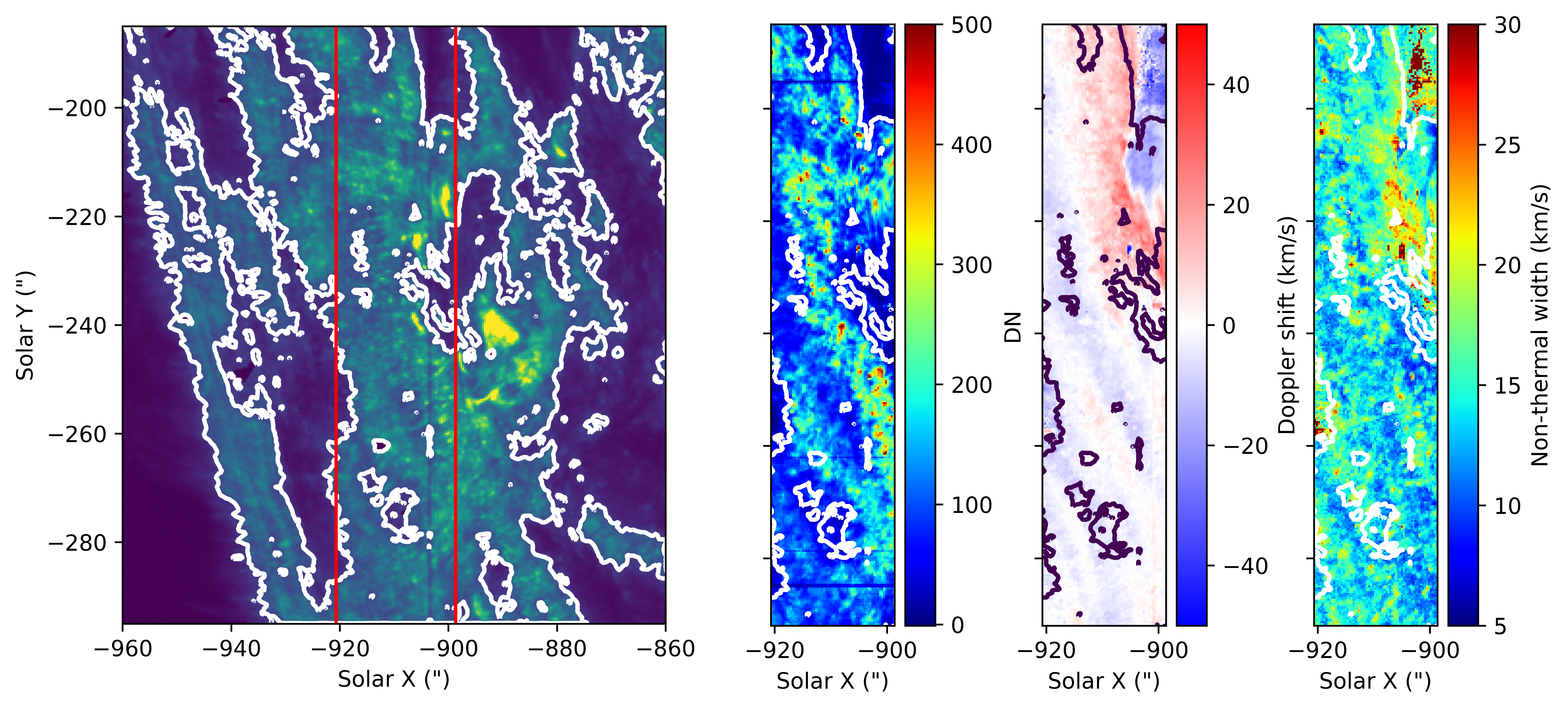}
	\caption{ From left to right, \iris\ SJI 1330 at 23:28:32, the total intensity, Doppler shift, and non-thermal width from \ion{Si}{4}~1339~\AA\ from fitting a single Gaussian. The contours correspond to the SJI 1330 at 100~DN. The vertical red lines in the left panel limit the region of the raster scan.}
\end{figure*}

We fit \ion{Si}{4}~1339~\AA\ profiles with a single Gaussian to extract the Doppler shifts and non-thermal widths (see Fig~\ref{fig:obsntw}). The non-thermal velocity is determined from the 1/e width of the single Gaussian fit by taking the square root of the square of the 1/e width after subtracting the squares of the instrumental and thermal contributions \citep[see,][]{Testa2016ApJ...827...99T}. Since the observation is near the limb, the non-thermal width contains mostly unresolved velocities that are expected to be almost perpendicular to the magnetic field. One can recognize spicules and moss in the SJI and \ion{Si}{4} line intensity map (left two panels). In plage and spicules, most non-thermal velocities range from $10$ to $25$~km~s$^{-1}$. Note that the exposure time is 2~s, so the assumption that the unresolved velocity comes from Alfvén waves suggests that they may have high frequencies (1~s or smaller) with amplitudes within $[15, 36]$~km~s$^{-1}$. We note that there is also plentiful observational evidence for Alfv\'en waves at lower frequencies \citep{De-Pontieu:2007bd,Okamoto:2011kx,McIntosh:2011fk,Tomczyk:2007vn}.

\section{Characterizing the ponderomotive acceleration and chemical fractionation in rMHD simulation: LTE vs NEQ}~\label{sec:smhd}

This section briefly describes the single fluid radiative MHD numerical model and plasma properties in the simulated chromosphere, including the ponderomotive force (calculated after the simulation was run), to understand the conditions under which a FIP effect could possibly occur. 

\subsection{Radiative MHD numerical model including NEQ ionization and ambipolar diffusion}

For the forward analysis, we used an already analyzed radiative MHD model computed with Bifrost \citep{Gudiksen:2011qy}, which includes non-equilibrium ionization for hydrogen and helium \citep{Leenaarts:2007sf,Golding:2016wq} as well as ambipolar diffusion \citep{Nobreg-Siverio:2020AA...633A..66N}. In short, this 2.5D numerical model mimics two plage regions connected by $\sim40$~Mm long loops and reveals features resembling type I and II spicules, low-lying loops, and other features. The simulation spans a vertical domain stretching from $\sim3$~Mm below the photosphere to $40$~Mm above into the corona with a non-uniform vertical grid size of $12$~km in the photosphere and chromosphere and $14$~km grid size in the horizontal axis. For further details about the numerical model, we refer to \citet{Martinez-Sykora:2020ApJ...889...95M}. We have chosen this model because it includes NEQ effects and produces type II spicules and the associated Alfvén waves \citep{Martinez-Sykora:2017sci}.

\subsection{Characterizing the simulated chromospheric properties for the chemical fractionation}

\begin{figure}
    \includegraphics[width=0.49\textwidth]{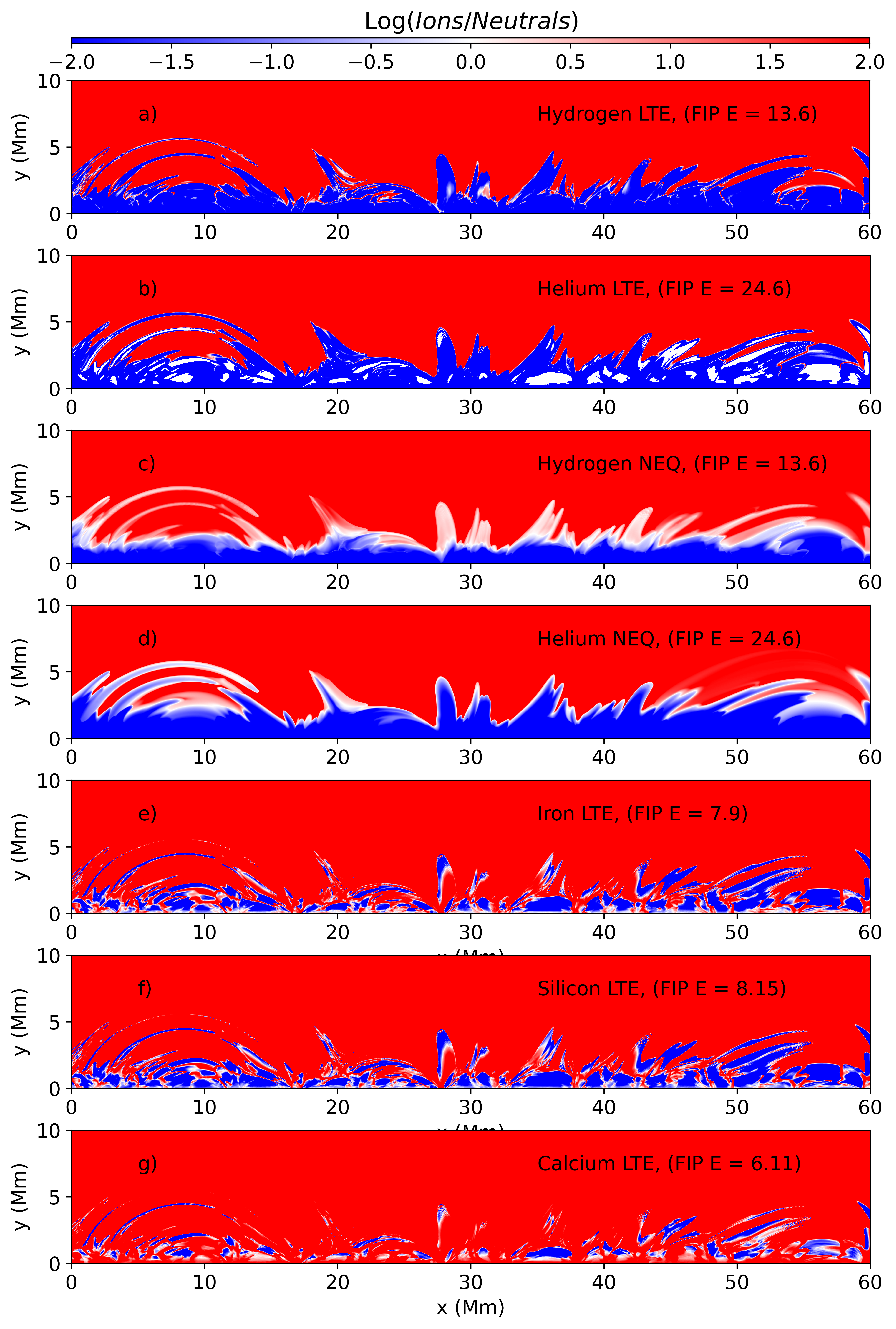}
	\caption{\label{fig:ionfrac} Ionization fractions are shown for various species with low and high FIP (hydrogen in panels a and c, helium in panels b and d, iron in panel e, silicon in panel f, and calcium in panel g). Panels a and b assume LTE, and panels c and d are in NEQ for hydrogen and helium. Note the color bar is in log-scale. Labels in the panels show the FIP energy in eV.}
\end{figure}

In this numerical model, we computed the ionization fraction for various species with both low (calcium, iron, silicon) and high (hydrogen and helium) FIP (Figure~\ref{fig:ionfrac}). We derive from the model the ionization fraction for hydrogen and helium in LTE (panels a and b) as well as in NEQ ionization (panels c and d). We assumed photospheric abundances \citep{Asplund:2009uq}, and statistical and thermal equilibrium for the ionization and density populations for neutrals and ions in the LTE cases. Low FIP species,  e.g., calcium, iron, and silicon, are highly ionized in a large fraction of the chromosphere in LTE, whereas high FIP species, including hydrogen and helium, are mostly neutral in the chromosphere in LTE. However, NEQ ionization effects significantly increase the ionization fraction for hydrogen (panel c) and helium (panel d) and lead to a larger height range where these species are partially ionized \citep{Leenaarts:2007sf,Golding:2016wq}. 

\begin{figure*}
    \includegraphics[width=0.97\textwidth]{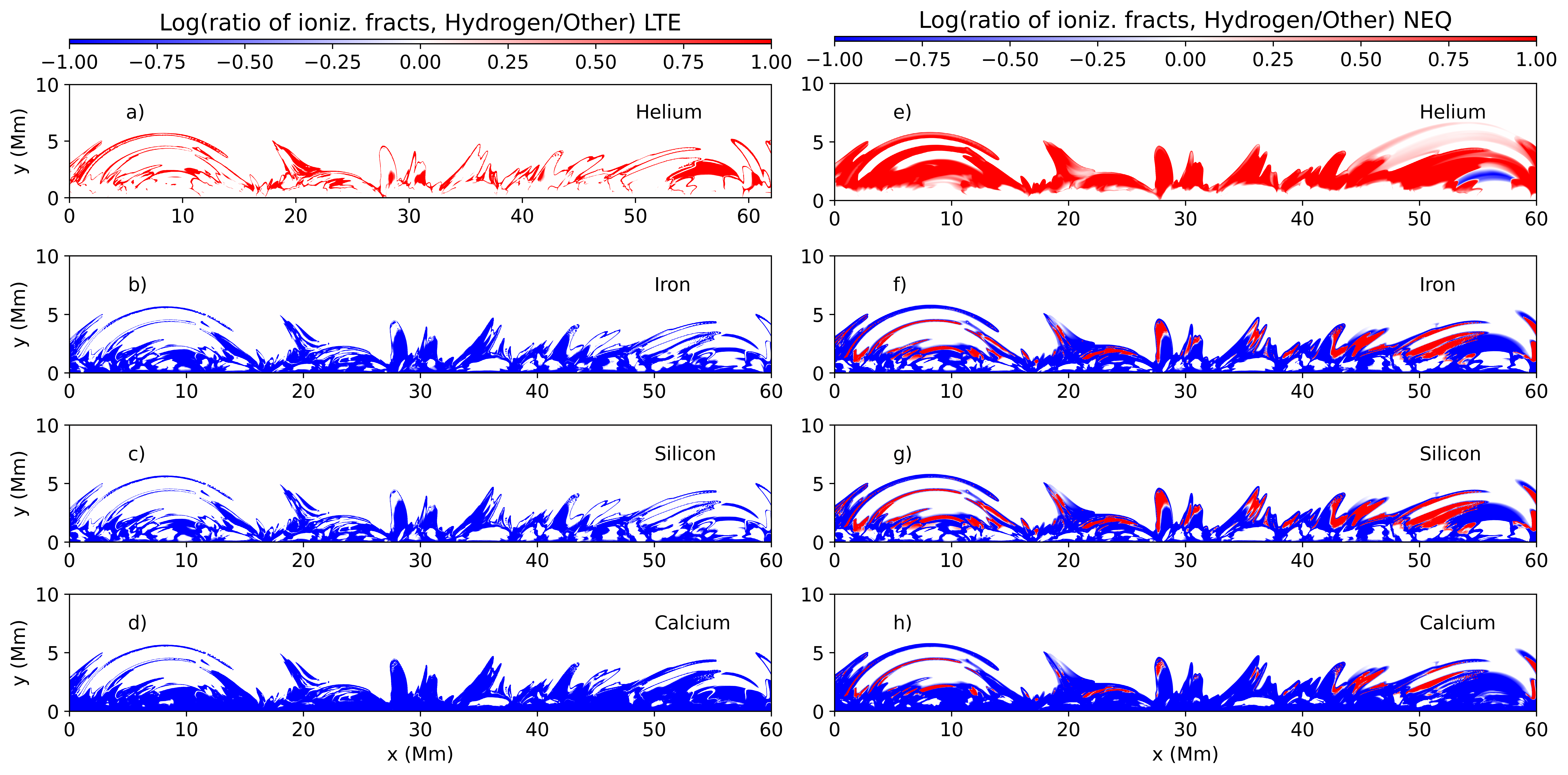}
	\caption{\label{fig:ionratio} The ratio of the ionization fraction of hydrogen with other species (helium, iron, silicon, and calcium from top to bottom) reveals where the relative chemical fractionation may happen. Red means hydrogen is ionized, whereas the other is neutral and blue vice-versa. The left column is in LTE, and the right column NEQ for hydrogen and helium.}
\end{figure*}

A comparison between hydrogen ionization fraction and any other species ionization fraction allows us to visualize the region of interest where the first ionization occurs for the various species (for low and high FIP) and chemical fractionation may occur. Figure~\ref{fig:ionratio} shows the ratio of hydrogen ionization fraction with the ionization fraction of any other species. We considered this comparison in LTE (left column) and NEQ ionization for hydrogen and helium (right column). The red or blue regions are regions of interest in terms of fractionation since in these regions the ionization degree between low and high FIP elements is significantly different (either higher or lower, respectively). If a physical process like the ponderomotive force were to act in these regions, it would potentially lead to relative enrichment (or depletion) of elements with different FIP. In addition, in NEQ, the ionization degree may be inverted in extended areas (red color), i.e., the hydrogen is ionized when the other species are neutral, e.g., panels f-i). 

\begin{figure*}
    \includegraphics[width=0.97\textwidth]{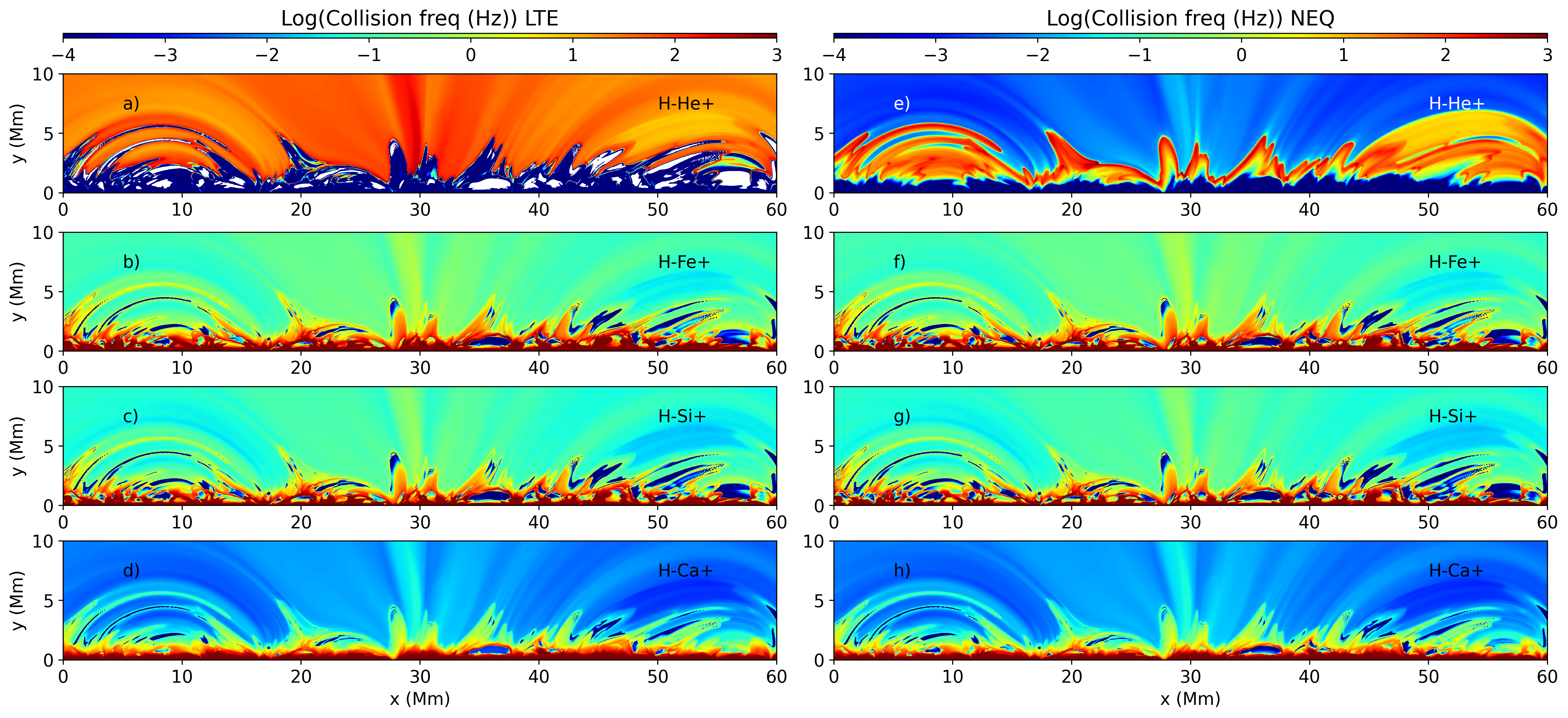}
	\caption{\label{fig:coll} Collision frequency between the neutral hydrogen and the other ionized species (helium, iron, silicon, and calcium from top to bottom) maps are shown in LTE (left column) and NEQ ionization (right column).}
\end{figure*}

However, we note that to have chemical fractionation, a different degree of ionization between the various species is not sufficient since: 1. a physical process needs to act on the various species (e.g., ponderomotive force) to accelerate the particles upward, 2. the dynamics of the various species need to be somewhat decoupled from the timescales involved in the acceleration (i.e., collisions between high and low FIP elements cannot dominate since these can couple the dynamics of the various elements). We compute the collisional rates between hydrogen, which is the most abundant fluid and will provide a good estimate of coupling, and other ionized species, again, considering LTE (left) and NEQ (right) in Figure~\ref{fig:coll}. The upper chromosphere and transition region are weakly collisional. Weakly collisional regions is where collisional timescales are a significant fraction of the wave period. Therefore, we consider weakly collisional is for collision frequencies lower than 10~Hz, and the driver of the chemical fractionation may occur in those locations. Note that in the simulation, the collision frequencies for ionized iron, silicon, and calcium are the same in NEQ and LTE because they do not depend on hydrogen number density. Their contribution to the momentum exchange nevertheless depends on the hydrogen number density \citep[e.g., see][]{Wargnier:2022ApJ...933..205W}.

\begin{figure*}
    \includegraphics[width=0.97\textwidth]{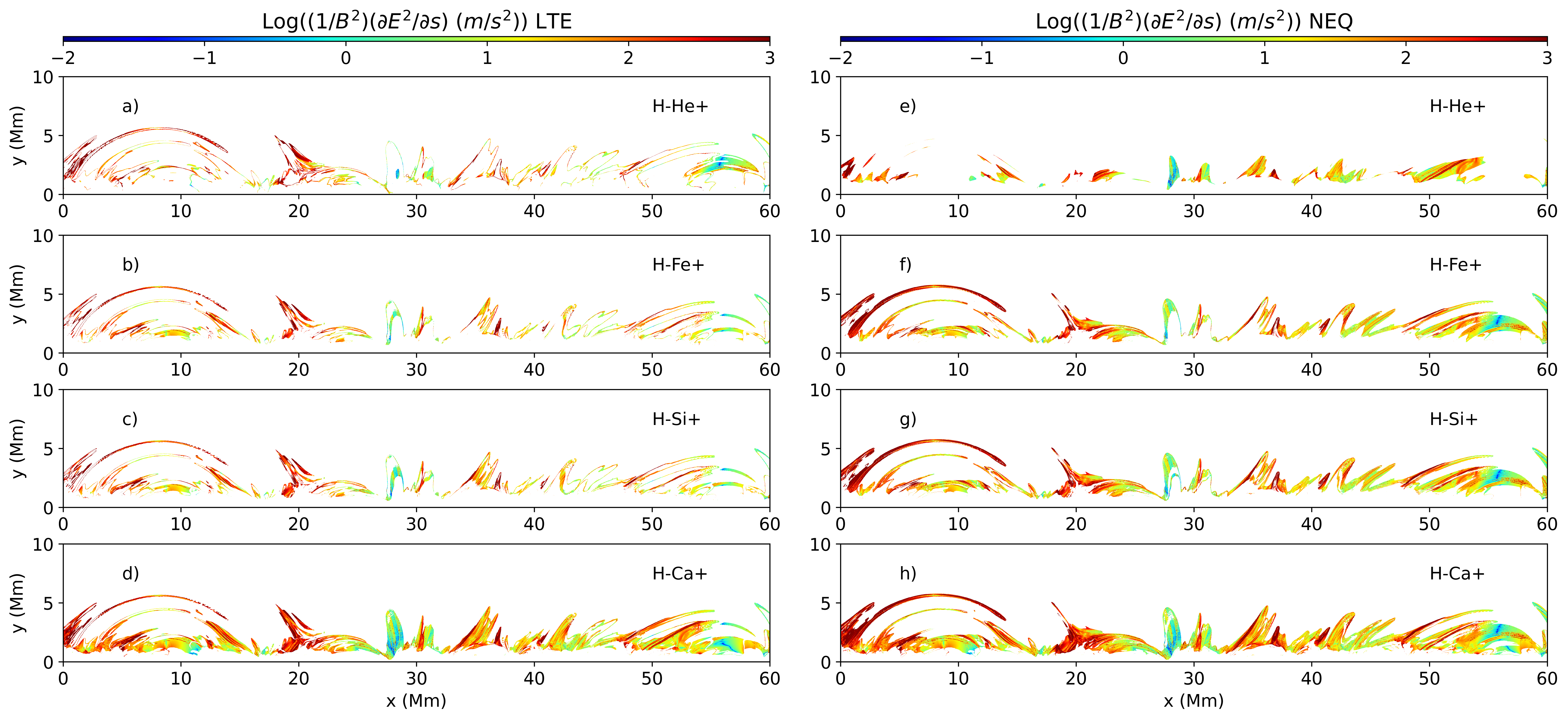}
	\caption{\label{fig:pondc} Maps of the electric field variations along the loop (Eq.~\ref{eq:pond3}). In these maps, we have masked out regions with high collision rates ($>10$~Hz) for each element (Figure~\ref{fig:coll}) as well as regions where they have the same ionization state as hydrogen (Figure~\ref{fig:ionratio}). The region of interest (non-masked region) is where the ponderomotive force may produce chemical fractionation. The distribution of the various panels is the same as in Figure~\ref{fig:coll}.}
\end{figure*}

A key aspect is where in the model the assumed physical process (ponderomotive force) is likely to be important. \citet{Dahlburg:2016ApJ...831..160D}  estimated the ponderomotive acceleration in their 3D model considering the electric field perturbations along the magnetic field. Similarly, we compute the electric field perturbation along the magnetic field as follows: 

\begin{eqnarray}
\delta_s E = \frac{1}{B^2}\frac{\partial E^2}{\partial s},~\label{eq:pond3}
\end{eqnarray}

\noindent where $s$ is along the loop. We mask regions with high collision rates ($\ge10$~Hz) for each element as well as regions where they have the same ionization state as hydrogen (Figure~\ref{fig:pondc}). By masking out these regions, we get the region of interest (the non-masked region), where the action of $|\delta_s E|$ may lead to chemical fractionation for each species. Different structures show different  $|\delta_s E|$, with type II spicules experiencing the largest values. Note that the region of interest changes depending on which species are considered. Those masks reveal the complexity of this problem due to the highly structured and dynamic atmosphere. 

\begin{figure*}
    \includegraphics[width=0.99\textwidth]{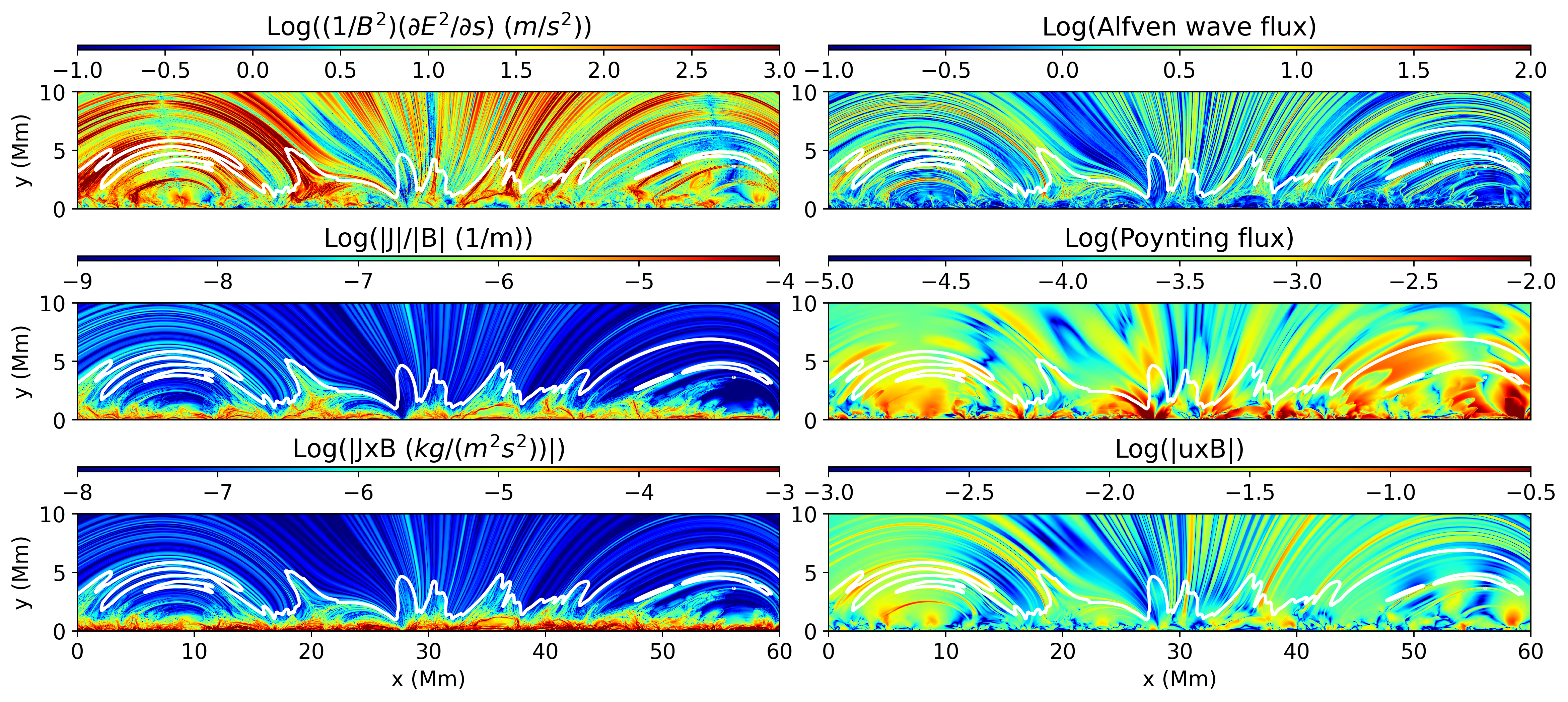}
	\caption{\label{fig:simvars} Various physical processes play a different role in different features. $\delta_s E$ following Eq~\ref{eq:pond3}, the Alfvén wave power, $|J|/|B|$, Poynting flux, Lorentz force and advection term of the electric field are shown from left to right and top to bottom, respectively. The solid line is at temperature $10^5$~K.}
\end{figure*}

Since the models in \citet{Dahlburg:2016ApJ...831..160D} were driven by waves, expression~\ref{eq:pond3} estimates the ponderomotive acceleration. This is not necessarily true in self-consistent radiative MHD since many other physical processes change the electric field and not only waves. To obtain a broader view, we show in Figure~\ref{fig:simvars}, in addition to $\delta_s E$, the Alfvén wave power \cite[by computing the velocity projections onto three characteristic directions, e.g.,][]{Khomenko_2011}; $|\vec{J}|/|\vec{B}|$, which reveals regions with strong tension or reconnection; the Poynting flux; the Lorentz force; and the advective part of the electric field $|\vec{u}\times \vec{B}|$. One can appreciate that there is not a single physical process that correlates or connects directly with $\delta_s E$ or Alfvén waves. Some locations with high $\delta_s E$ are associated with several other physical processes. Still, regions with spicules show either the largest values of $|\vec{J}|/|\vec{B}|$ or Lorentz force ($x=20$, 35 or 40~Mm) as well as $\delta_s E$. On top of this, those regions generate the strongest Alfvén wave power that penetrates the corona. We find that regions with large $|\vec{J}|/|\vec{B}|$ are associated with spicules or low-lying loops interacting with small and large-scale fields. The field lines connected to large $|\vec{J}|/|\vec{B}|$, and hence large $\delta_s E$ in the chromosphere, shows large Alfvén wave amplitude in the transition region and corona. Note that these regions are connected to regions with enhanced magnetic field networks or plage regions. These then are the regions that seem to harbor conditions that are favorable for the ponderomotive force to act. Observationally these are often regions that show enhanced non-thermal line broadening, as we show in Section~\ref{sec:irisanalysis}. In principle, high-frequency Alfvén waves can produce large non-thermal velocities of transition region lines. This is because these waves are often unresolved at the temporal and spatial resolution of current instruments (see Section~\ref{sec:irisanalysis}). 

\begin{figure*}
    \includegraphics[width=0.97\textwidth]{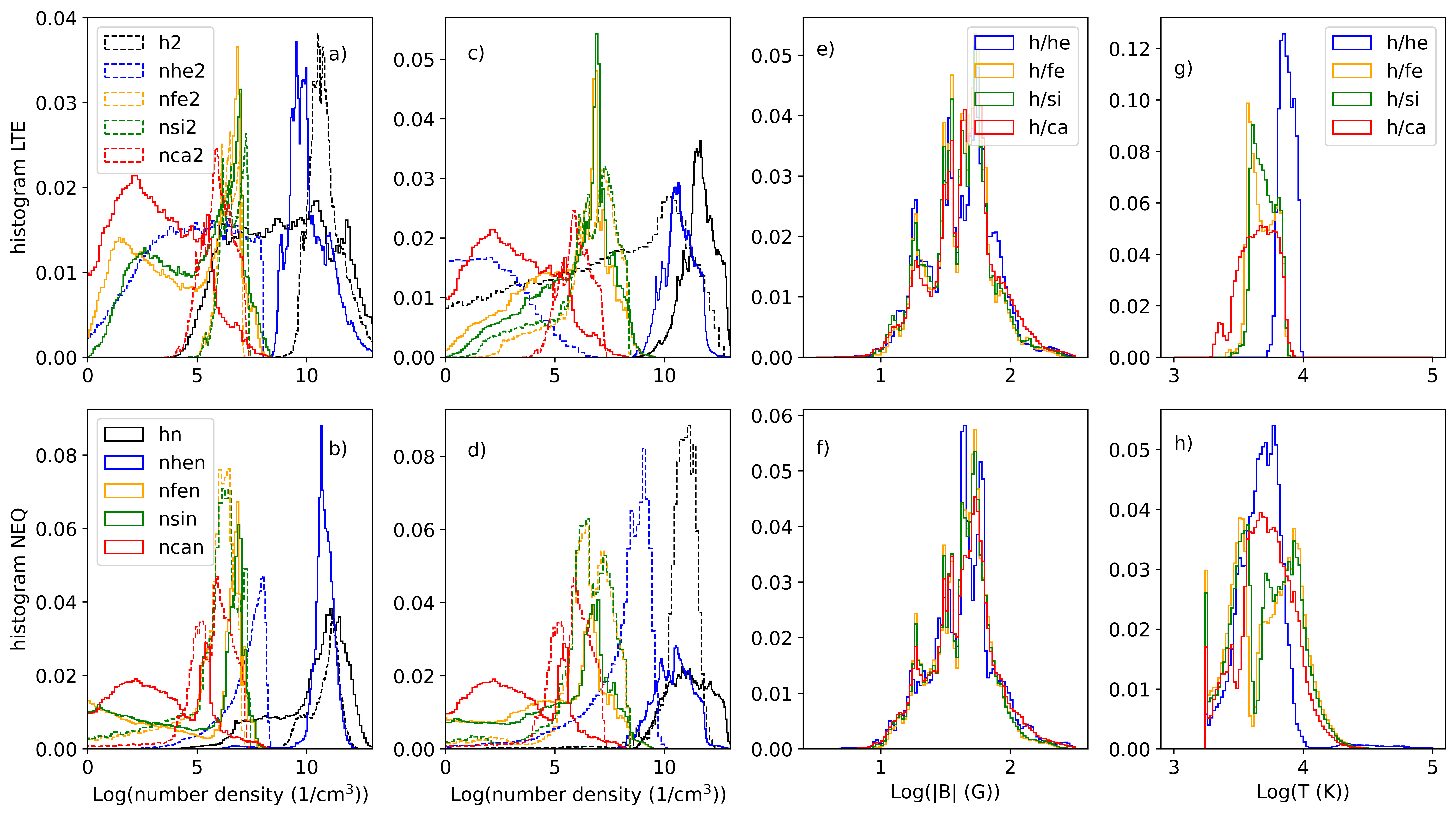}
	\caption{\label{fig:pophist} Histogram of the density population for neutrals (solid, panels a-d) and ions (dashed, panels a-d), magnetic field (third column) and temperature (right column) within the regions of interest (non-masked out regions) in Figure~\ref{fig:pondc}, for each species  (panels a and b, and e-h). Panels c and d are for the region of interest of H-Ca. LTE case is in the top row and NEQ in the bottom row.}
\end{figure*}

In order to initialize our MFMS simulations, we compute within the regions of interest (the non-masked regions in Figure~\ref{fig:pondc}), the populations of each species (four left panels of Figure~\ref{fig:pophist}), magnetic field (panels e and f) and temperature (panels g and h). The top row is under LTE conditions, and the bottom row is for NEQ. Panels a, b, e-h are within the regions of interest for each species, and panels c and d are for the region of interest of H-Ca (panels d and h in Figure~\ref{fig:pondc}). Note that the  regions of interest are where the chemical fractionation may happen since it is relatively weakly collisional and the ionization degree differs between high and low FIP elements.  To inspire the initial conditions of our MFMS simulations (see below), we use the region of interest of H-Ca, because it covers the largest region where chemical fractionation may happen. From the models it appears that the field strength ranges from $\sim10$~G to $\sim 120$~G and the median is around 30~G in those regions where chemical fractionation may occur. For the temperature, there is a clear difference between LTE and NEQ scenarios; the former has a narrower temperature range ($\log( T (K))=[3.4,4]$) than in NEQ ($\log( T (K))=[3.2,4.3]$) and both roughly peaking at similar temperature ($\log( T (K))\sim 3.7$), but for LTE this varies more for each species. 

\begin{figure}
    \includegraphics[width=0.49\textwidth]{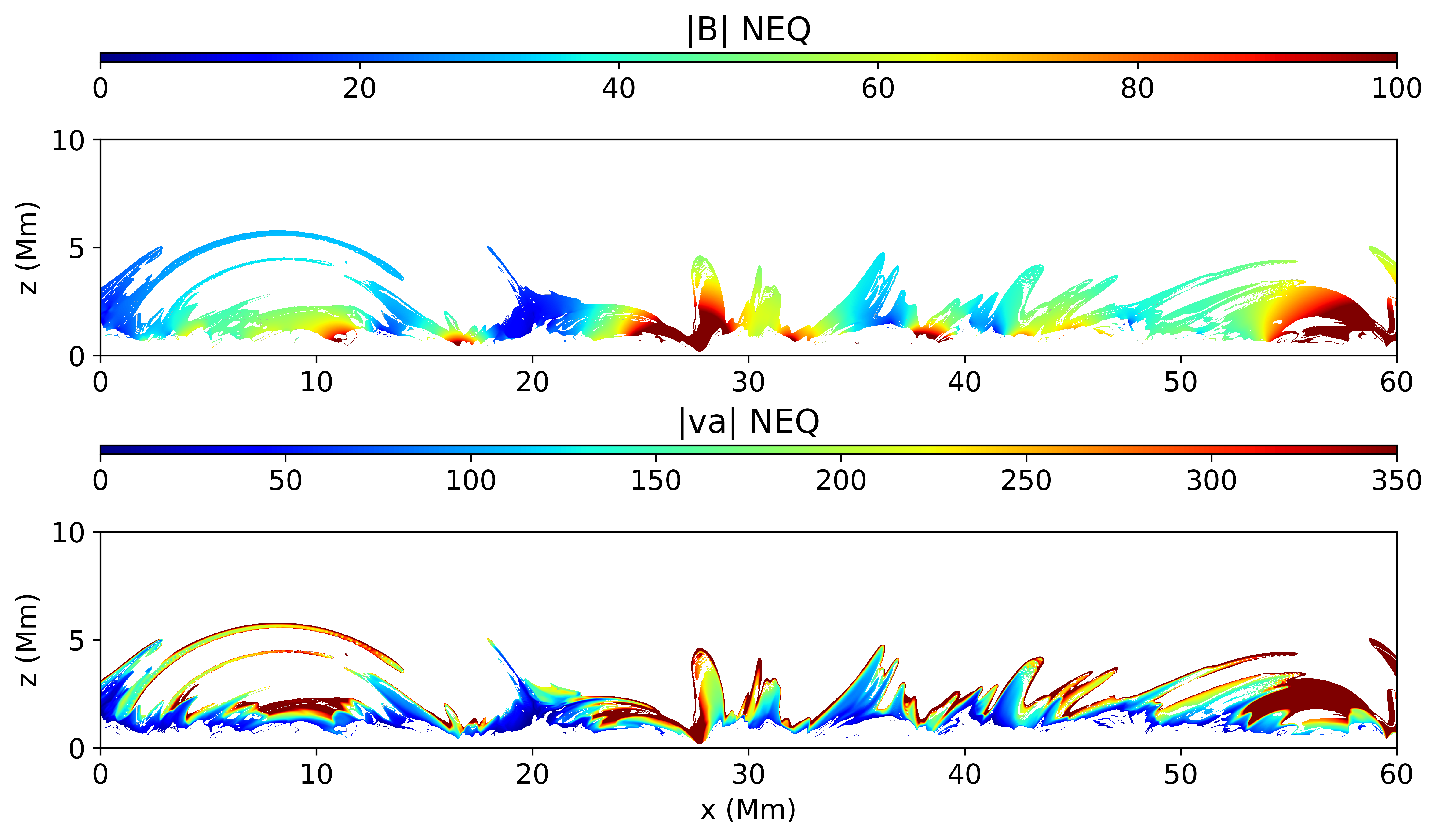}
	\caption{\label{fig:modb_map} Magnetic field strength (top) and Alfvén speed (bottom) maps, where we applied the same mask as in panel h in Figure~\ref{fig:pondc}, reveal the complexity of the expansion or canopies of the magnetic field topology within the regions of interest.}
\end{figure}

For the following study (Section~\ref{sec:mfms}) and to limit the parameter range of the multi-fluid simulations (which are computationally expensive), we restrict the variation of the magnetic field along the loop, and initialize the densities as constant in space (Section~\ref{sec:theo}). This means that we consider 1D loops that expand or constrict with height (and/or direction of wave propagation). One would expect that on the Sun the degree of expansion with height of the field would depend on the type of region, e.g., coronal holes, quiet sun, plage, active regions, etc. We compute the magnetic field strength and Alfvén speed  in the region of interest shown in panel h of Figure~\ref{fig:pondc} from the 2.5D numerical model and find that in our plage simulation the configuration is much more complex than a simple expansion with height, and this would need to be taken into account in the future (Figure~\ref{fig:modb_map}). For now, inspired by the model field topology, we will parameterize the expansion of the magnetic field as listed in Table~\ref{tab:1ds}. The complex magnetic field expansion or canopies have roughly a variation of the magnetic field of $\Delta B \sim \pm 20$~G. 

\section{Multi-fluid and multi-species simulations of the ponderomotive force}~\label{sec:mfms}

In this second part of this manuscript, we introduce the multi-fluid multi-species (MFMS) numerical code Ebysus, including a short description of the physics included and the numerical scheme (Section~\ref{sec:ebcode}). As a continuation, we describe the setup of the parametric study of the various numerical models (Section \ref{sec:ebsetup}) and, finally, the results of the MFMS models (Section~\ref{sec:ebres}). 

\subsection{The MFMS numerical Ebysus code}~\label{sec:ebcode}

The Ebysus code is the first of its kind to treat each excited/ionized level for each desired species as a separate fluid. The multi-fluid equations in SI implemented in the code are as follows: 

\begin{widetext}

\begin{eqnarray}
\frac{\partial \rho_\mathrm{aIE}}{\partial t} + \nabla \cdot\left( \rho_\mathrm{aIE}\,  {\vec u_\mathrm{aIE}}\right) & = &  %\nonumber \\ 
%&& 
\sum_{I'E',a} m_\mathrm{aIE}(n_\mathrm{aI'E'}\Gamma^{ion}_\mathrm{aI'E'IE}-n_\mathrm{aIE}\Gamma^{rec}_\mathrm{aIEI'E'}) \label{eq:conti}  \\
\frac{\partial (\rho_\mathrm{a\hat{I}E} {\vec u_\mathrm{a\hat{I}E}}) }{\partial t}+ \nabla \cdot (\rho_\mathrm{a\hat{I}E} {\vec u_\mathrm{a\hat{I}E}} \otimes {\vec u_\mathrm{a\hat{I}E}} -\hat{\tau}_\mathrm{a\hat{I}E}+P_\mathrm{a\hat{I}E}\identity)  & = & %\nonumber \\
%&& 
\rho_\mathrm{a\hat{I}E} {\vec g}+ n_\mathrm{a\hat{I}E} q_\mathrm{a\hat{I}} \left({\vec E} + {\vec u_\mathrm{a\hat{I}E}} \times {\vec B} \right) + \nonumber \\
&& 
\sum_{\hat{I'}E',a}(\Gamma_\mathrm{a\hat{I'}E'\hat{I}E}^{ion}m_\mathrm{a\hat{I}E}\vec{u}_\mathrm{a0E'}-\Gamma_\mathrm{a\hat{I}EI'E'}^{rec}m_\mathrm{a\hat{I}E}\vec{u}_\mathrm{a\hat{I}E}) %\nonumber \\
%&& 
+\sum_\mathrm{a'I'E'}\vec{R}_\mathrm{a\hat{I}E}^{\mathrm{a\hat{I}Ea'I'E'}} \label{eq:spmoma1}\\
\frac{\partial (\rho_\mathrm{a0E} {\vec u}_\mathrm{a0E}) }{\partial t}+ \nabla \cdot (\rho {\vec u}_\mathrm{a0E}\otimes {\vec u}_\mathrm{a0E} -\hat{\tau}_\mathrm{a0E}+P_\mathrm{a0E}\identity)  & = & %\nonumber \\
%&&  
\rho_\mathrm{a0E} {\vec g} +\sum_\mathrm{a'I'E'}\vec{R}_\mathrm{a0E}^{\mathrm{a0Ea'I'E'}}- \nonumber \\ 
&& 
\sum_{\mathrm{I'E',a}}\left(\Gamma_\mathrm{aI'E'}^{ion}m_\mathrm{a0E}\vec{u}_\mathrm{a0E}+\Gamma_\mathrm{a0E}^{rec}m_\mathrm{aI'E'}\vec{u}_\mathrm{aI'E'}\right) 
\label{eq:spmomc1}\\
\frac{\partial{e_\mathrm{aIE}}}{\partial{t}}   +\nabla\cdot \left(e_\mathrm{aIE}\vec{u}_\mathrm{aIE}\right) +P_\mathrm{aIE}\nabla\cdot\vec{u}_\mathrm{aIE}=   &&%\nonumber \\
%&&  
Q^{visc}_\mathrm{aIE} + Q^{ir}_\mathrm{aIE} + Q^{\mathrm{aIEe}}_\mathrm{aIE} + \sum_\mathrm{a'I'E'}^{\mathrm{a'I'E'}\neq \mathrm{aIE}} Q_\mathrm{aIE}^{\mathrm{aIEa'I'E'}} \label{eq:enei5} \\ 
\frac{\partial{e_{e}}}{\partial{t}} + \nabla\cdot \left(e_{e}\vec{u_{e}}\right) + P_{e}\nabla\cdot\vec{u_{e}}= Q^{visc}_e  &+&Q^{ir}_e + %\nonumber \\
%&& 
\sum_\mathrm{a'I'E'} Q_{e}^{e\mathrm{a'I'E'}} + Q_e^\mathrm{spitz} \label{eq:enee5} 
\end{eqnarray} 

\end{widetext}

\noindent where the ionization states are referred as $\mathrm{I}$, i.e., $\mathrm{I}=0$ denotes neutrals and $\hat{\mathrm{I}} = \mathrm{I} \geq 1$ ions. The excited levels are marked with $\mathrm{E}$ and the identity of the chemical species (or molecules) is indicated by $\mathrm{a}$. Consequently, each set of particles in a given micro-state will be described with $\mathrm{aIE}$. For electrons, the notation $\mathrm{aIE}$ is reduced to just $\mathrm{e}$. For simplicity, $\sum_\mathrm{a'}$ is the sum over all the species $\mathrm{a'}$, $\sum_\mathrm{I',a}$ is the sum over all ionization levels, including neutrals, for a given species $\mathrm{a}$ and $\sum_{\mathrm{E}',\mathrm{aI}}$ is the sum over all the excited levels for a given ionized species $\mathrm{aI}$. For clarity, we define $\sum_\mathrm{a'I'E'} = \sum_\mathrm{a'}\sum_\mathrm{I',a'}\sum_\mathrm{E',a'I'}$, and  $\sum_{I'E',a} = \sum_\mathrm{I',a}\sum_\mathrm{E',aI'}$. The mass density is $\rho_\mathrm{aIE}=m_\mathrm{aIE}\, n_\mathrm{aIE}$, where ${\vec u_\mathrm{aIE}}$, $n_\mathrm{aIE}$ and $m_\mathrm{aIE}$ are the velocity, number density and  particle mass for a given micro-state. $\Gamma^{rec}_\mathrm{aIEI'E'}$, and $\Gamma^{ion}_\mathrm{aI'E'IE}$ are the transition rate coefficients between levels $I'E'$ and $IE$ due to recombination or de-excitation, and ionization or excitation, respectively. $q_\mathrm{aI'E'IE}$, $P_\mathrm{aI'E'IE}$, and $\hat{\tau_\mathrm{aI'E'IE}}$ are the ion charge, gas pressure, and viscous tensor for a specific species.  ${\vec g}$, ${\vec E}$, and ${\vec B}$ are gravity acceleration, and electric and magnetic field, respectively. $\vec{R}_\mathrm{aIE}^{\mathrm{aIEa'I'E'}}$ is the momentum exchange where $\mathrm{aIE} \neq \mathrm{a'I'E'}$. For the collision integrals, we consider neutral-neutral, ion-neutral including charge exchange, or Maxwell molecular collisions, and Coulomb collisions between ions following 
\citet{Bruno:2010PhPl...17k2315B} 
\citep[see][for further details on the collisions and cross-sections]{Wargnier:2022ApJ...933..205W}. 
The momentum exchange can then be expressed as follows: 

\begin{eqnarray}
\vec{R}_{\mathrm{aIE}}^{\mathrm{aIEa'I'E'}} = m_{\mathrm{aIE}}\, n_{\alpha}\, \nu_{\mathrm{aIEa'I'E'}} ( \vec{u}_{\mathrm{a'I'E'}} - \vec{u}_{\mathrm{aIEa}}),
\end{eqnarray}

\noindent where  $\nu_{\mathrm{aIEa'I'E'}}$ is the collision frequency. Note that, to guarantee the conservation of the total momentum, $\vec{R}_{\mathrm{aIE}}^{\mathrm{aIEa'I'E'}} = - \vec{R}_{\mathrm{a'I'E'}}^{\mathrm{a'I'E'aIE}}$.  $Q^{visc}$, $Q^{ir}_\mathrm{aIE}$, $Q_{\mathrm{aIE}}^{\mathrm{aIEa'I'E'}}$, and  $Q_\mathrm{spitz}$ are the viscous heating due to hyper-diffusion, ionization and recombination, momentum exchange, and Spitzer thermal conduction heating/cooling terms, respectively. $Q_{\mathrm{aIE}}^{\mathrm{aIEa'I'E'}}$ includes the thermalization between the various fluids. 

We ignore electron inertia and its time variation. So, in Ebysus, we consider the magnetic induction equation:

\begin{eqnarray}
\frac{\partial{\vec{B}}}{\partial{t}} &=& - \nabla\times\vec{E}  \\ \nonumber
&=&\nabla\times \left(\vec{u_e} \times \vec{B} - \frac{\nabla P_e}{n_\mathrm{e} q_\mathrm{e}} + \frac{\sum_\mathrm{aIE}\vec{R}_{\mathrm{e}}^{\mathrm{eaIE}}}{n_\mathrm{e} q_\mathrm{e}}\right)
\end{eqnarray}

\noindent the second and third terms are the Bierman battery and the ohmic diffusion, respectively, and we ignored the ionization recombination contribution to the electric field. Since electrons move so fast and their mass is negligible we assume quasi-neutrality: $n_\mathrm{e}=\sum_\mathrm{aIE} n_\mathrm{aIE}\, Z_\mathrm{aI}$ where $Z_\mathrm{aI}$ is the ionized state, and we neglect their inertia. Since several ionized species are considered, the electron velocity, thanks to the assumption of quasi-neutrality, reads as follows: 

\begin{eqnarray}
\vec{u}_{e} = \left(\sum_\mathrm{a\hat{I}E} \frac{n_\mathrm{a\hat{I}E} q_\mathrm{a\hat{I}E} \vec{u}_\mathrm{a\hat{I}E}}{n_\mathrm{e}\, q_\mathrm{e}}\right) - \frac{\vec{J}}{q_\mathrm{e}\, n_\mathrm{e}}, ~\label{eq:vele}
\end{eqnarray}

\noindent and $\vec{J} = (\nabla \times \vec{B})/\mu_0$. We refer to the total ion velocity as ${\vec u}_c=\sum_\mathrm{a\hat{I}E}{\vec u}_\mathrm{aIE}$ and for neutrals as  ${\vec u}_n=\sum_\mathrm{a0E}{\vec u}_\mathrm{a0E}$.

Ebysus has inherited the numerical methods from Bifrost \citep{Gudiksen:2011qy}, and the first results of this code can be found in \citet{Martinez-Sykora:2020ApJ...889...95M}. In short, the numerical mesh is defined in a staggered Cartesian box. The spatial derivatives and interpolation of the variables are 6th and 5th-order polynomials, respectively. To advance in time, we can select which source terms are advanced explicitly or implicitly using either Lie or Additive operator splittings. For the explicit part, we stepped forward in time using the modified explicit third-order predictor-corrector Hyman method \citep{Hyman:1979acmp.proc..313H}, or we can also choose a 3rd-order Runge-Kutta method. The implicit part can be solved using different methods, some of which are mentioned below and further details will be described in a dedicated manuscript for the code. Finally, the numerical noise is suppressed using a high-order artificial diffusion. 

Due to the numerical stiffness, the ionization, recombination, excitation and de-excitation, momentum exchange, and the ohmic terms have been added and are solved by applying a Newton-Raphson implicit method, or the fifth-order implicit Runge Kutta method known as Radau IIA method \citep{HAIRER199993,Hairer1996}. 

For this research, we neglect ionization and recombination, gravity, and the heating or cooling sources from radiation, ohmic diffusion, and thermal conduction.  These assumptions allow us to ``isolate" and limit the parameter range of the considered dependencies of the ponderomotive force on Alfvén waves.

\subsection{Initial and boundary conditions} \label{sec:ebsetup}

We investigate the effects of MFMS interactions and LTE vs NEQ ionization on the ponderomotive force due to Alfvén waves in a parametric study for chromospheric conditions. The density and temperature are taken from the median of the histograms in Figure~\ref{fig:pophist} assuming the region of interest from panels d (LTE) or h (NEQ) in Figure~\ref{fig:pondc} and listed in Table~\ref{tab:pop}. We consider one LTE case. For computational reasons, we had to select higher ion densities to be able to run this LTE scenario for hydrogen and helium. This is because the very different range of values of number density for the various fluids leads to computational complications. The temperature is constant with $1.6\times10^4$~K. Finally, the magnetic field follows a quarter of a sinusoidal profile which varies within the values listed in the sixth column of Table~\ref{tab:1ds} where the first number is at the bottom and the last is at the top.  The range for the Alfvén speed is listed in the sixth column. The values and range of the magnetic field and Alfvén speed are similar to the parameter range within the region of interest shown in Figure~\ref{fig:modb_map} except near the transition region where the Alfvén speed variations with height can become much larger than the range covered here. Because we assume constant densities, the variation of the Alfv\'en speed in the box has the opposite sign to that of the magnetic field. This is contrary to what would be expected on the Sun since the density stratification leads to an increase in Alfv\'en speed with height, despite the decrease in magnetic field strength expected from an expanding flux tube. This should be considered when we refer to the magnetic field variation along the loop as the ``expansion" or ``constriction" of field lines in loops. The variation of the Alfv'en speed along the loop is expected to play a significant role in ponderomotive force and is thus not fully considered here, given our constant density assumption. 

The numerical domain in all 1.5D numerical experiments is along the loop, which for the MFMS models will be the $z$-axis, and covers the range $z=[0,3]$~Mm with a uniform grid of 1000 points, which is roughly the length  where the fractionation may occur in panel j) of Figure~\ref{fig:pondc}. The grid spacing has been chosen to resolve spatially and temporally the Alfvén waves for any simulation. Note that $z=0$ is not the photosphere as in the previous section but the range $z=[0,3]$~Mm covers a subregion within the chromosphere. In all of our simulations, waves are launched from z=0~Mm, and propagate towards positive values of z. As the wave propagates it will either encounter an increase of magnetic field strength (i.e., a constricting flux tube), or a decrease of magnetic field strength (i.e., an expanding flux tube).

\begin{table}
	\centering
	\caption{Density number of the various fluids. From left to right: the species, and the number density for the neutral and ionized fluid in cm$^{-3}$}\label{tab:pop}	
	\begin{tabular}{|c|c|c|}
		\hline
		Species & Neutrals (NEQ, LTE) & Ions (NEQ, LTE)  \\ \hline  \hline
		H & $10^{9.8}$, $10^{11.1}$  & $10^{10.7}$, $10^{8.5}$\\ \hline
        He & $10^{9.7}$, $10^{10.6}$ & $10^{8.3}$, $10^{4.8}$\\ \hline
		Fe & $10^{4.6}$, $10^{5.25}$ & $10^{5.6}$, $10^{6.4}$ \\ \hline
		Si  &$10^{4.7}$, $10^{5.7}$ & $10^{5.4}$, $10^{6.2}$ \\ \hline
		Ca & $10^{2.7}$, $10^{2.5}$ & $10^{5.3}$, $10^{5.9}$ \\ \hline
	\end{tabular}
\end{table}

The boundaries are open at the ``top" of the domain. At the ``bottom," we drive an Alfvén wave along the component $x$ of the magnetic field with different frequencies and amplitudes listed in the two rightmost columns of Table~\ref{tab:1ds}. We put quotation marks around the words ``top" and ``bottom" because our numerical experiments do not include gravity. The bottom to top direction is really the direction of wave propagation. Note that the frequencies are inspired from the observational constraints in Section~\ref{sec:iris} and the selected amplitudes are a fraction of the derived velocity amplitudes from the non-thermal velocity assuming that not all of the broadening is caused by Alfvén waves. Another reason to not use large amplitudes is to avoid non-linearity effects in this initial study.  Columns 2-5 list which physical processes are included in each of the simulations. 

\begin{table*}
	\centering
	\caption{List of  numerical simulations. From left to right: list the name, physical processes included or not, magnetic field configuration, Alfvén speed range, and frequency and amplitude of the Alfvén wave driver. In the fifth column, the first value of the $B$ range is the bottom boundary, and the second is the top, so $[70, 50]$~G is a loop that its magnetic field strength decreases with $z$, i.e., direction of wave propagation.}\label{tab:1ds}	
	\begin{tabular}{|c|c|c|c|c|c|c|c|}
		\hline
		Name & Ohm & Coll & Hall & $B$ range (G) & $v_A$ (km~s$^{-1}$) & Fr. (Hz) & $B$ Amp. wave (G) \\ \hline  \hline
		NH\_B7050\_F1\_D1 & No & No & No & $[70,50]$ & $[290,210]$ & 1   & 0.1 \\ \hline
		NC\_B7050\_F1\_D1 & No & No & Yes & $[70,50]$ & $[290,210]$ & 1   & 0.1 \\ \hline
		NC\_B5070\_F1\_D1 & No & No & Yes &  $[50,70]$ & $[210,290]$ & 1   & 0.1 \\ \hline
		C\_B7050\_F1\_D1  & No & Yes & Yes & $[70,50]$ & $[290,210]$ & 1   & 0.1 \\ \hline
		C\_B5070\_F1\_D1  & No & Yes & Yes & $[50,70]$ &  $[210,290]$ & 1   & 0.1 \\ \hline
    	A\_B8040\_F1\_D1  & Yes & Yes & Yes & $[80,40]$ & $[330,170]$ & 1   & 0.1 \\ \hline
		A\_B6555\_F1\_D1  & Yes & Yes & Yes & $[60,55]$ & $[270,220]$ & 1   & 0.1 \\ \hline
		A\_B5070\_F1\_D1  & Yes & Yes & Yes & $[50,70]$ & $[210,290]$ & 1   & 0.1 \\ \hline
		A\_B7050\_F10\_D1 & Yes & Yes & Yes & $[70,50]$ & $[290,210]$ & 10 & 0.1 \\ \hline
		A\_B7050\_F2\_D1 & Yes & Yes & Yes & $[70,50]$ & $[290,210]$ & 2 & 0.1 \\ \hline
		A\_B7050\_F01\_D1 & Yes & Yes & Yes & $[70,50]$ & $[290,210]$ & 0.2 & 0.1 \\ \hline
		A\_B7050\_F1\_D5 & Yes & Yes & Yes & $[70,50]$ & $[290,210]$ & 1   & 0.5 \\ \hline 
		A\_B7050\_F1\_D1  & Yes & Yes & Yes & $[70,50]$ & $[290,210]$ & 1   & 0.1 \\ \hline
		A\_B7050\_F1\_D05 & Yes & Yes & Yes & $[70,50]$ & $[290,210]$ & 1   & 0.05 \\ \hline
		A\_B7050\_F1\_D01 & Yes & Yes & Yes & $[70,50]$ & $[290,210]$ & 1   & 0.01 \\ \hline
		LTE\_B5070\_F1\_D1  & Yes & Yes & Yes & $[70,50]$ & $[225,160]$ & 1   & 0.1 \\ \hline
	\end{tabular}
\end{table*}

\subsection{Results of the MFMS models} \label{sec:ebres}

In Section~\ref{sec:theo}, we emphasize the challenging work of selecting the proper assumptions to choose the appropriate ponderomotive force. In contrast, the Ebysus code self-consistently solves the multi-fluid equations for various species without considering any of those particular assumptions. As we will show, the assumptions required for calculating the ponderomotive forces can be very limiting, which can lead to an overestimation or misjudgement of their impact on the plasma dynamics. 

The ponderomotive force depends on the variation of the electric field along the loop, while the electric field associated with Alfvén waves depends on the wave energy. Therefore, first, we investigate the velocity and magnetic field variation as a function of space. This will help to understand the variation of the electric field with the distance along the loop (i.e., magnetic field lines). These gradients are ultimately responsible for the acceleration along the loop. 

\begin{figure*}
    \includegraphics[width=0.97\textwidth]{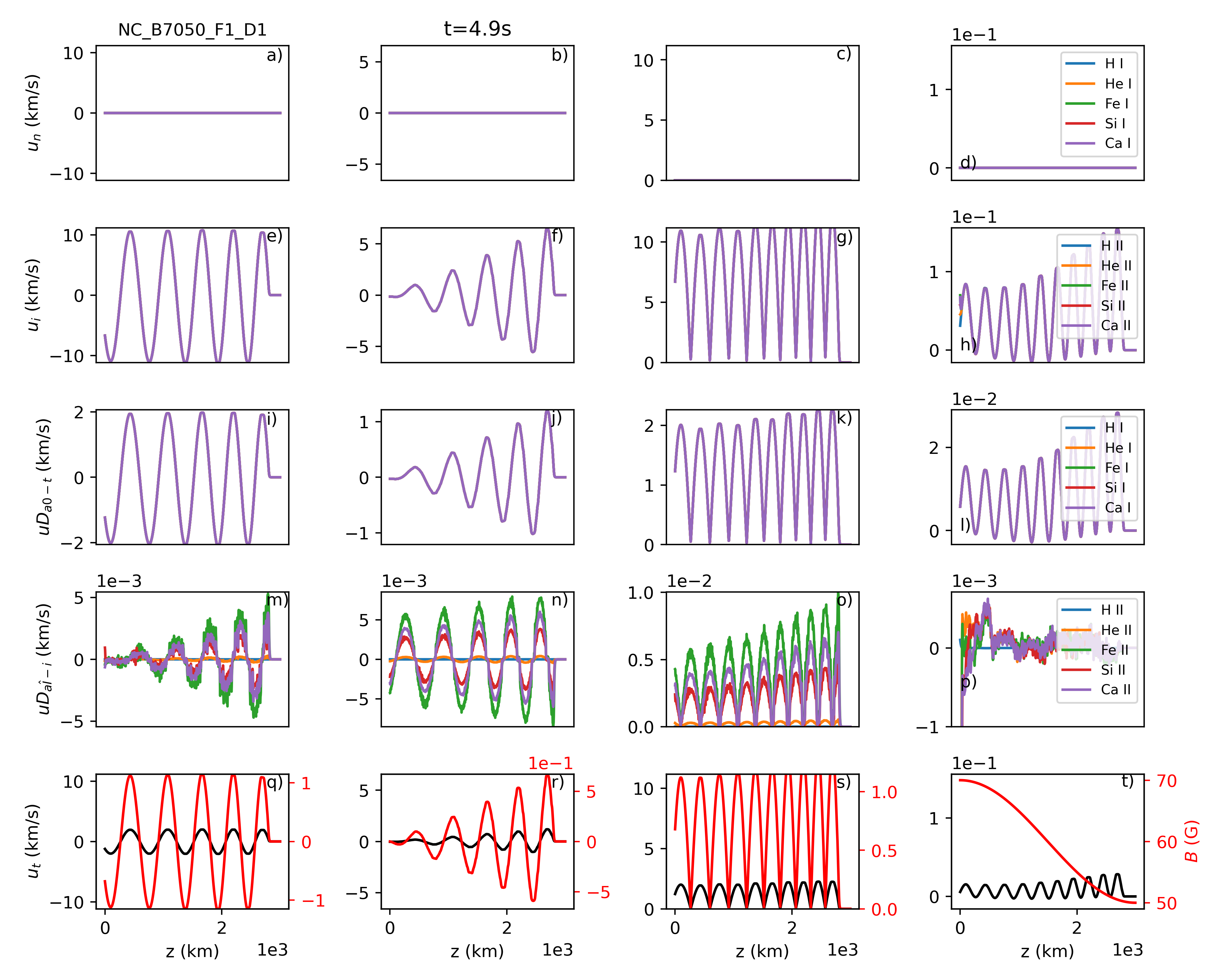}
	\caption{\label{fig:eb_space_nocol} Spatial variation at t=5~s of velocity for all the neutral species (top row), ion species (second row), velocity drift for each neutral fluid with respect the total velocity of all fluids (third row), velocity drift for individual ion-fluids with respect to the combined fluid of ions (fourth row). In the bottom row, we show the magnetic field (red) and single fluid velocity, i.e., the total velocity of all fluids (black). For all rows, we show the $x$ component, the $y$ component, the absolute value of the component perpendicular to the guide field, and the $z$ component, from left to right columns, respectively. This figure is for simulation NC\_B7050\_F1\_D1, i.e., it excludes collisions. In this case the field strength decreases in the direction of wave propagation (e.g., for a wave propagating upwards in a loop that expands with height).} 
\end{figure*}

Let us focus first on a case without collisions, e.g., NC\_B7050\_F1\_D1. Figure~\ref{fig:eb_space_nocol} shows various physical parameters as a function of space once the wave has propagated through most of the domain. Since there are no collisions, neutral fluids do not experience any variation in space and time as a result of an Alfvén wave. However, it is interesting to see a rotation of the ion velocities (second row) and magnetic field (bottom row in red) perpendicular to the guided field, as shown with the two perpendicular components ($x$, and $y$ in the first and second columns). The rotation is due to the ion coupling between the various ionized fluids as reported by \citep{Martinez-Sykora:2020ApJ...889...95M} and not because of the Hall term. The equivalent simulation without the Hall term (NH\_B7050\_F1\_D1) also has this rotation (not shown here). The ion coupling leads to a velocity drift between different ions (fourth row). Furthermore, due to the expansion of the magnetic field, i.e., a magnetic field strength decrease with $z$ and similarly with the Alfvén speed, there is an upflow in the ion velocity (right column), which can produce a chemical fractionation depending on the ionization fraction of each species. It is also interesting to see the amplitude of the ion velocities (panels e-h) and the perpendicular component of magnetic field (panels q-s) increase. The wavelength decreases with $z$, i.e., with the decrease of the magnetic field (panel t, red), This is expected from a propagating Alfvén wave in a medium where the Alfvén speed decreases with $z$. Note that the Alfvén speed follows the magnetic field strength since the density and temperature is constant. 

\begin{figure*}
    \includegraphics[width=0.97\textwidth]{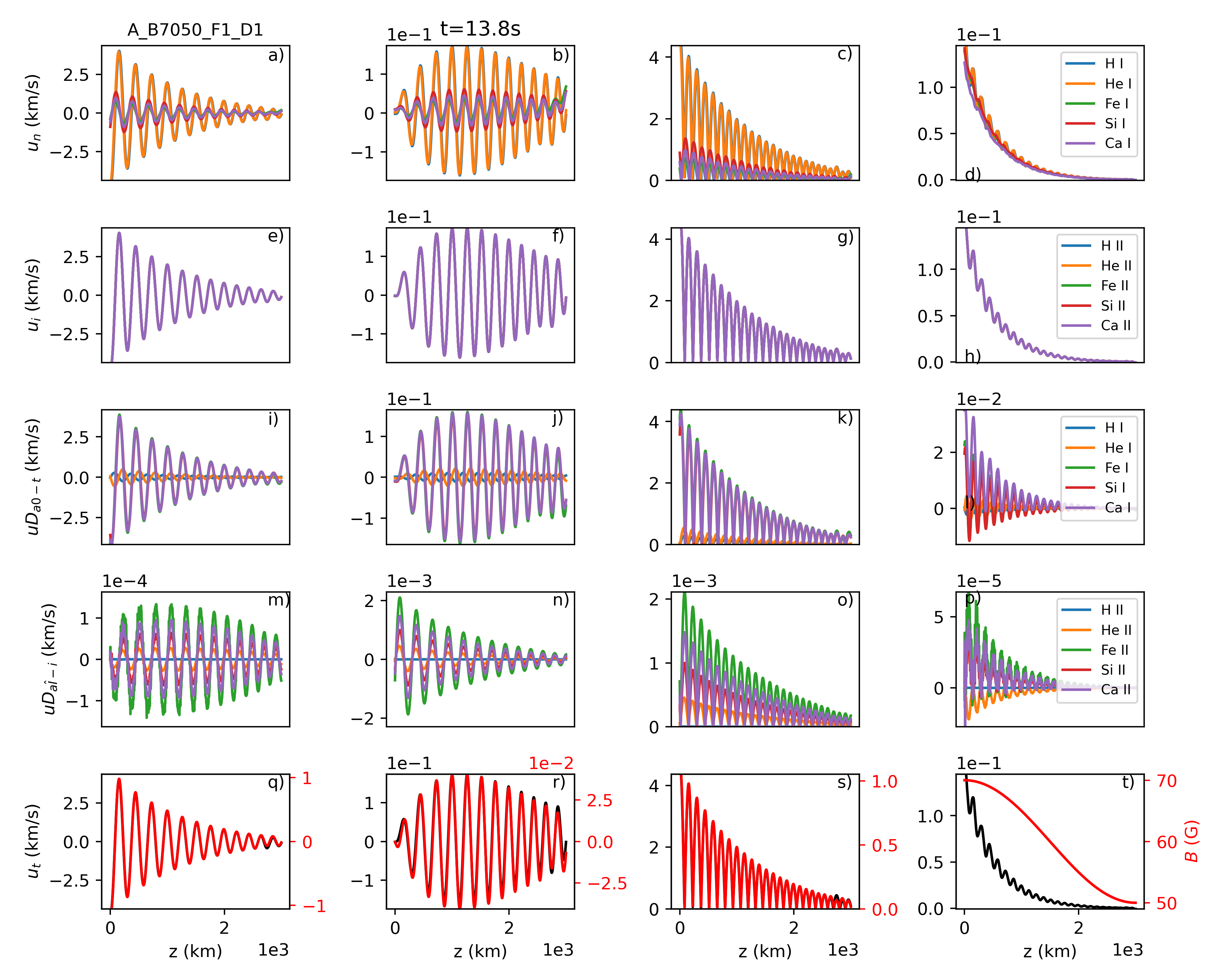}
	\caption{\label{fig:eb_space} Same layout as Figure~\ref{fig:eb_space_nocol} for simulation A\_B7050\_F1\_D1, i.e., with collisions and for a decreasing field strength in the direction of the wave propagation (i.e., for a wave  propagating upwards in a loop that expands with height).}
\end{figure*}

This picture changes drastically when collisions are included, e.g., A\_B7050\_F1\_D1 (Figure~\ref{fig:eb_space}). Neutrals are being dragged by the ions (top row), and the velocity patterns for each neutral fluid are shifted spatially because of a phase shift, which depends on the collision frequency with other ions. Still, neutrals are not completely coupled with ions, and a small velocity drift is present (third row). As a result, the wave power, i.e., the velocity perpendicular to the magnetic field, drops while it propagates along the $z$-axis, i.e., dissipation occurs. Interestingly, vertical flows are much larger than the collisionless scenario shown above. Like for the amplitudes, the vertical velocities drop drastically with $z$ in contrast to the previous case. This is because of the presence of collisions. Neutrals show relatively large velocity drifts, which result from a vertical velocity variation with $z$  due to collisional coupling, which does not occur in the collisionless scenario shown previously. Consequently, the multi-fluid interactions lead to a mix of wave phases between the various fluids, producing dissipation and a large contribution to the ponderomotive force (as shown below) compared to the collisionless case shown above.

\begin{figure*}
    \includegraphics[width=0.97\textwidth]{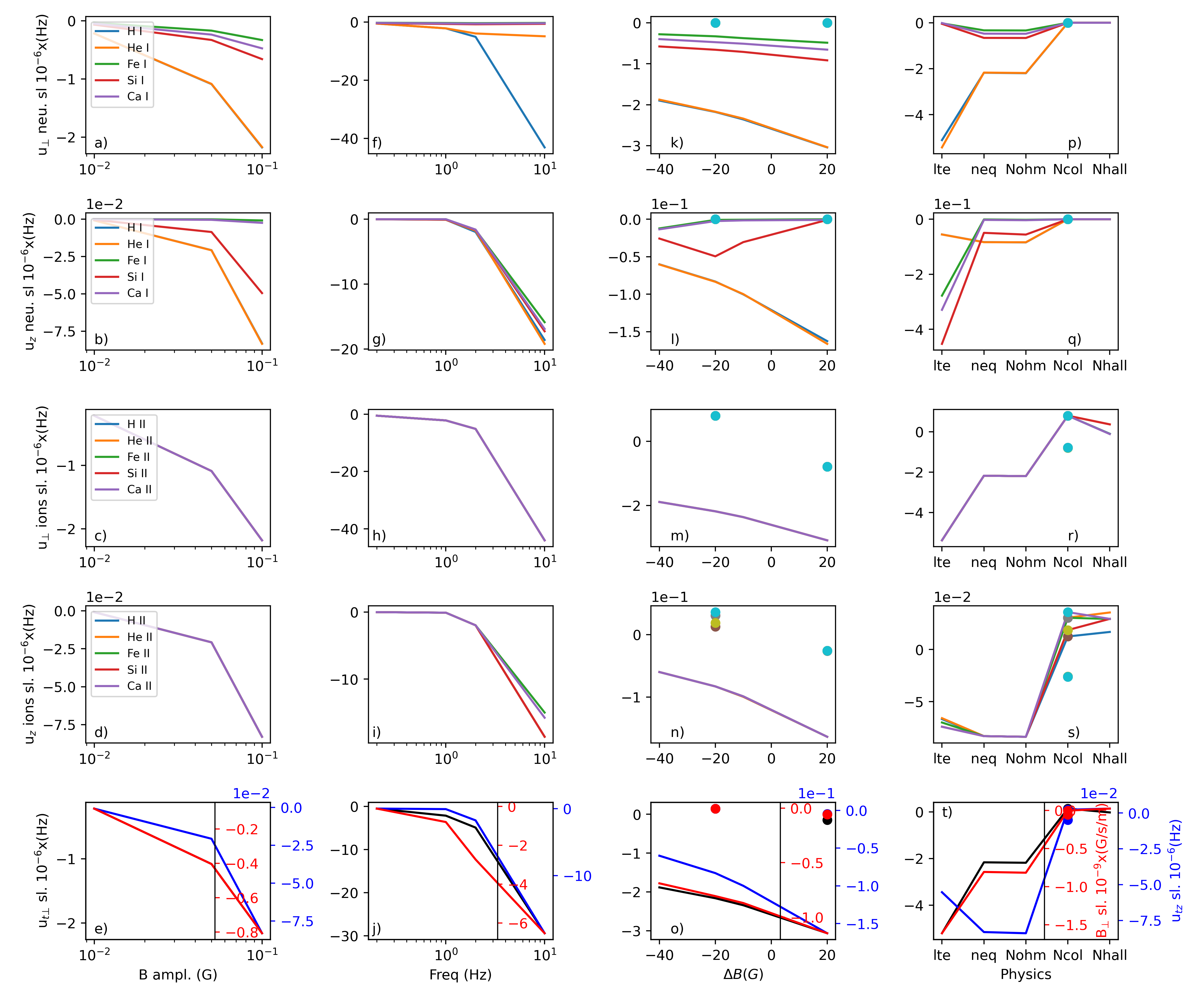}
	\caption{\label{fig:slopes_z} The median of the spatial gradient with $z$ of the wave-associated velocities that are perpendicular to the field (first and third rows) and longitudinal (second and fourth rows)  for neutrals (top two rows) and ions (third and fourth rows). We plot these as a function of amplitude and frequency of the Alfvén wave driver, variation of the guide magnetic field, and inclusion of various physical approaches, respectively 
 from left to right. The last row shows the spatial gradient with $z$ for wave-associated velocities (for all fluids combined) that are perpendicular to the field (black) and longitudinal (blue). We also show in the bottom row the spatial variation of the perpendicular component of the magnetic field (red). The collisionless cases have been added with dots in the two right columns.}
\end{figure*}

We now study the role of the parameters listed in Table~\ref{tab:1ds} on the variation of the wave properties as a function of $z$ (i.e., the direction of wave propagation, not necessarily height). To this end we calculate (and show in Fig.~\ref{fig:slopes_z}), the gradient with $z$ of the peaks of the oscillations for the various types of velocities: perpendicular to the field (first and third rows, for neutrals and ions respectively), along the field (second and fourth rows, for neutrals and ions respectively), and the total velocity of all fluids combined (fifth row). As we can see, most of the cases we studied show a negative gradient with $z$, i.e., a decrease with increasing $z$, except for a few cases without collisions. An increase in the amplitude (first column) or frequency (second column) of the driver leads to stronger gradients of velocities with $z$ and greater drifts for neutrals (top panels) in both perpendicular and longitudinal motions. The slopes of the various variables as a function of the background magnetic field variation with $z$ provide a fascinating result. For all the collisional cases, the slopes are negative, regardless of whether the loop is expanding ($\Delta B < 0$) or constricting ($\Delta B > 0$) in the direction of wave propagation. The collisionless cases (shown with dots) clearly have a different sign of the slopes (panels m and n), i.e., it is positive for $\Delta B < 0$ (expanding) and negative for $\Delta B > 0$ (constricting). For the collisional cases, the gradient with $z$  for all variables increases with increasing magnetic field expansion (decreasing $\Delta B$ for all negative $\Delta B$ cases). We will go into further details below. Finally, one can appreciate that the important physical processes are the presence of collisions and, in more minor roles, the ohmic diffusion and the Hall term (right column). In the right column, we added the impact of the LTE initial setup. Due to the differences in the populations of the various species, the interactions between species and the gradients of the electric field, hence the ponderomotive force, change dramatically compared to the NEQ case. In LTE, the neutral density is higher than the ion density which results in a stronger dissipation of the Alfvén wave and coupling between ion and neutral fluids. Here again, we added dots for the non-collisional cases for different magnetic field configurations for comparison. 

\begin{figure*}
    \includegraphics[width=0.97\textwidth]{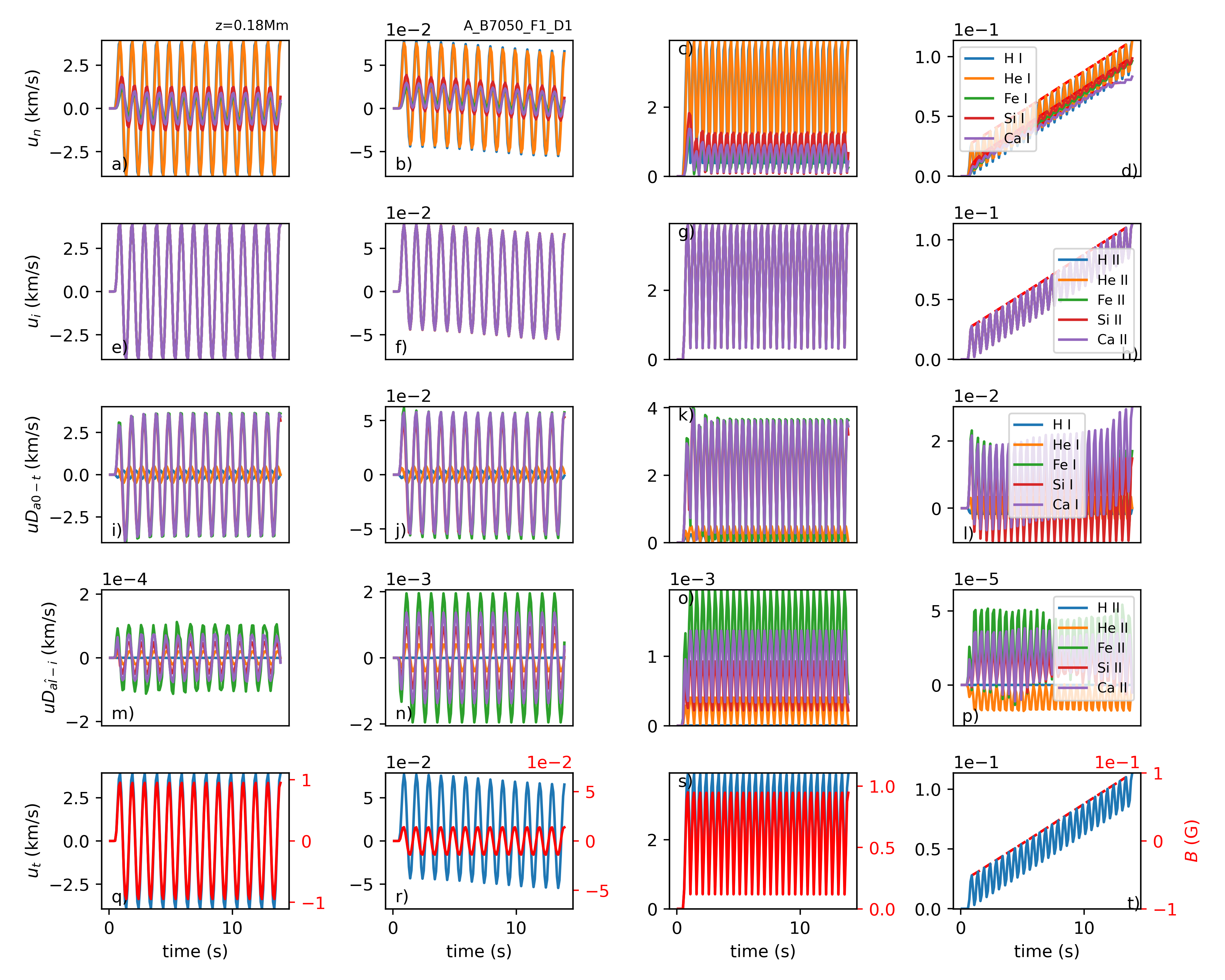}
	\caption{\label{fig:eb_time} Temporal variation at $z=0.18$~Mm of velocity for all the neutral species (top row), ion species (second row), velocity drift for each neutral fluid with respect to the total velocity of all fluids (third row), velocity drift for individual ion-fluids with respect to the combined ion-fluid (fourth row). In the bottom row, we show the magnetic field (red) and single fluid velocity, i.e., the total velocity of all fluids (black). For all rows, we show the $x$ component, the $y$ component, the absolute value of the component perpendicular to the guide field, and the $z$ component, from left to right columns, respectively. This figure is for simulation  A\_B7050\_F1\_D1, i.e., with collisions and a decreasing magnetic field strength with $z$.}
\end{figure*}

In order to study the variation of the acceleration for different fluids as a function of the physical parameters, first, we visualize the same variables as in Figure~\ref{fig:eb_space} (which are as a function of space for a given time), but now as a function of time at $z=0.18$~Mm. Note that there is no variation in the wave amplitude with time (Figure~\ref{fig:eb_time}). Therefore, the electric field has no dependence on time. Consequently, the second term in the parenthesis of Eq.~\ref{eq:pond1} is expected to be negligible. 
%This means that the assumptions  Abraham and Miller forces should not apply here and are hard to consider in the solar atmosphere (see also Section~\ref{sec:smhd}). 
Neutral fluids, as mentioned above, experience a different phase speed due to the differences in collisional frequencies. We note that neutral fluids experience more significant differences in vertical velocities than ions because the latter are magnetized and move with the magnetic field. A similar effect can be seen for the gradients with $z$, as shown in Figure~\ref{fig:eb_space}. Again, one can appreciate that there is a large phase mixing between the different types of fluids, which leads to the ponderomotive force, as we show below.

\begin{figure*}
    \includegraphics[width=0.97\textwidth]{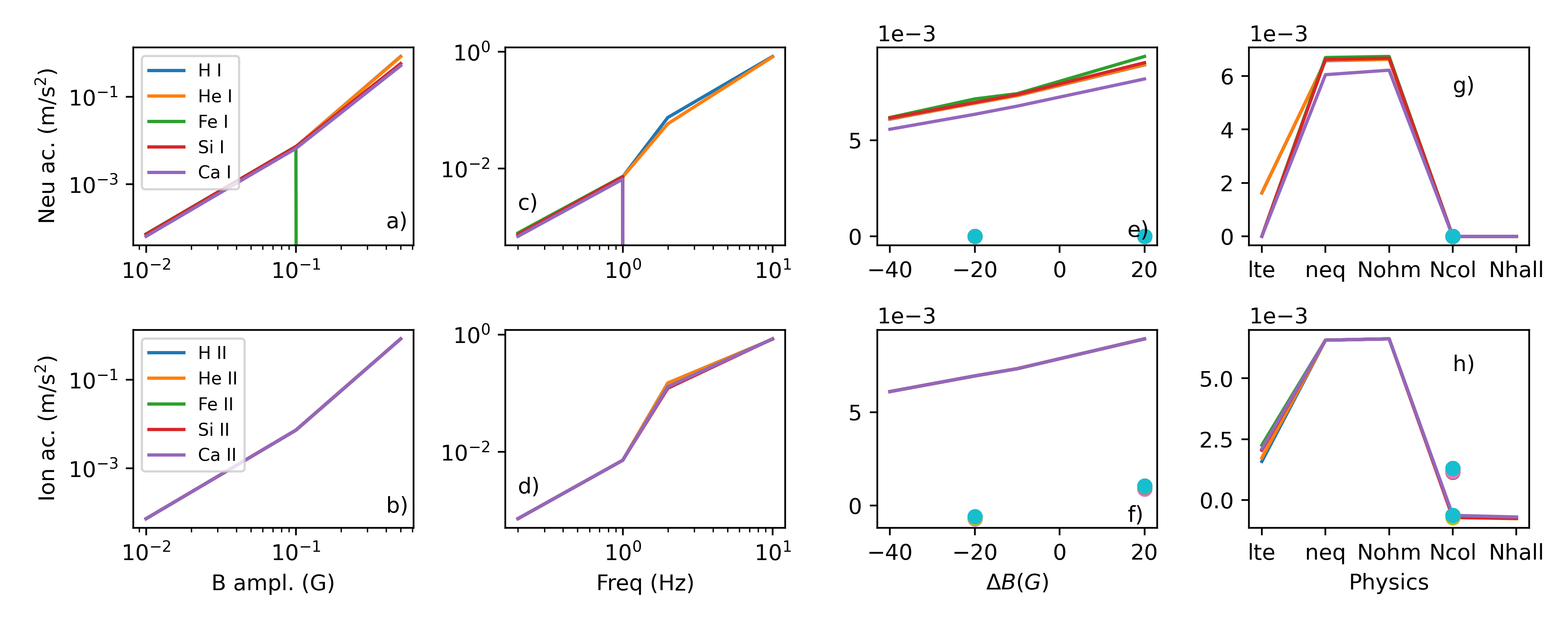}
	\caption{\label{fig:slopes_t} The median of the vertical acceleration at $z=0.18$~Mm for neutrals top(row) and ions (bottom row) as a function of Alfvén wave amplitude and frequency, the guided magnetic field variation, and physics are shown from left to right.}
\end{figure*}

We now calculate in Figure~\ref{fig:slopes_t} the gradients with time of the local maximum velocities in the $z$ direction (right column) of neutrals (top) and ions (bottom) shown with red dashed lines, similarly as for Figure~\ref{fig:slopes_z} but in time. This figure shows the estimates of the acceleration (in the direction of wave propagation) of the various fluids. The comparison of these accelerations, for each physical process, initial plasma properties, or imposed driver, allows us to further understand the action of the ponderomotive force associated with Alfvén waves in the chromosphere for various fluids (Figure~\ref{fig:slopes_t}). %We calculated the equivalent to Figure~\ref{fig:slopes_z} for the temporal variation of the vertical component of the velocities, i.e., accelerations. 
The acceleration increases with the amplitude and frequency of the driver. We remind the reader that the selected frequencies are inspired by \iris\ observations (Section~\ref{sec:iris}). It is very intriguing that the acceleration increases with frequency whereas Eqs.~\ref{eq:pond1}-\ref{eq:pond1c} are proportional to $1/\omega^2$. As shown in the third column, in the presence of collisions, there is always an acceleration in the direction of the wave propagation, no matter if the flux tube is expanding or constricting. There is some decrease in the acceleration with an increase in the expansion of the magnetic field, i.e., values of decreasing $\Delta B$ for the cases where $\Delta B$ is negative. %in accordance with the collisionless cases (dots). 
If we assume that a loop expands with height, and assume a collisional case, the plasma is accelerated in the same direction as the propagating waves. 
However, in the collisionless cases, if the loop constricts ($\Delta B > 0$) plasma is accelerated in the direction of wave propagation, while, if the loop expands ($\Delta B < 0$), plasma is accelerated in the direction opposite of the wave propagation. If we assume that a loop expands with height, and assume no collisions, this would mean that plasma is accelerated downward for upward propagating waves, and downward for downward propagating waves. It appears that, for the collisionless case, the sign disagrees with equations \ref{eq:pond1b}-\ref{eq:pond1c}, but agrees with the sign of equation \ref{eq:pond1}. In any case, the presence of collisions leads to a major qualitative and quantitative difference. 

Finally, the large accelerations result from collisions which lead to a vital phase mixing between the fluids, as shown in the right column in Figure~\ref{fig:slopes_t}. Note also the large difference between the LTE and NEQ cases. In LTE, the number density of neutrals is very large and collisions are high. Consequently, waves are damped within a very short traveled distance and the upflows and acceleration is highly localized closer to the source of the wave ($z\sim0$). 

We notice that for higher wave frequencies or amplitude, collisions dissipate the wave faster. Therefore, the accelerations in those cases are highly localized closer to the driver. Further away from the driver (which is at $z=0$), the spatial gradients of the variables and acceleration are close to zero. This is similar, as mentioned above, to the LTE case. In the latter case this due to the strong coupling with neutrals. 

%\begin{figure*}
%    \includegraphics[width=0.97\textwidth]{time_evol_fit_ac_iz850.png}
%	\caption{\label{fig:eb_time} Time plots at z=0.1Mm of velocity for all the neutral species (top row), ion species (second row), velocity drift for neutrals (third row), velocity drift for ions respect to the ion-fluid (fourth row), magnetic field (red) and single fluid velocity (blue,bottom row) are shown for component $x$, $y$, perpendicular to the guide field, and $z$, from left to right columns, respectively. This figure is for simulation  A\_B5070\_F1\_D1.}
%\end{figure*}

%\begin{figure*}
%    \includegraphics[width=0.97\textwidth]{slope_all_final_ac.png}
%	\caption{\label{fig:slopes_t} Median of the vertical acceleration at z=0.1Mm for neutrals top(row) and ions (bottom row) as a function of amplitude, and frequency of the Alfvén wave driver, guide magnetic field variation, and physics are shown from left to right.}
%\end{figure*}

In order to understand the cause of the vertical acceleration as a function of the various parameters, we compute (and show in Fig.~\ref{fig:eperp}), as a function of $z$, the electric field perpendicular to the magnetic field (top row), $- dE^2_{\bot}/dz$ (again calculated based on peaks associated with the wave) as it quantifies the ponderomotive acceleration (middle row), and the Alfvén speed (bottom row). Note that in our simulations, we find a negligible value for $dE^2/dt$, so we only consider the spatial gradient for $E^2$ in our proxy for the ponderomotive force. The dependence of the acceleration on the parameters shown in Figure~\ref{fig:slopes_t} can be understood by investigating the panel that shows $- dE^2_{\bot}/dz$ in the middle row of Figure~\ref{fig:eperp}: 

\begin{figure*}
    \includegraphics[width=0.97\textwidth]{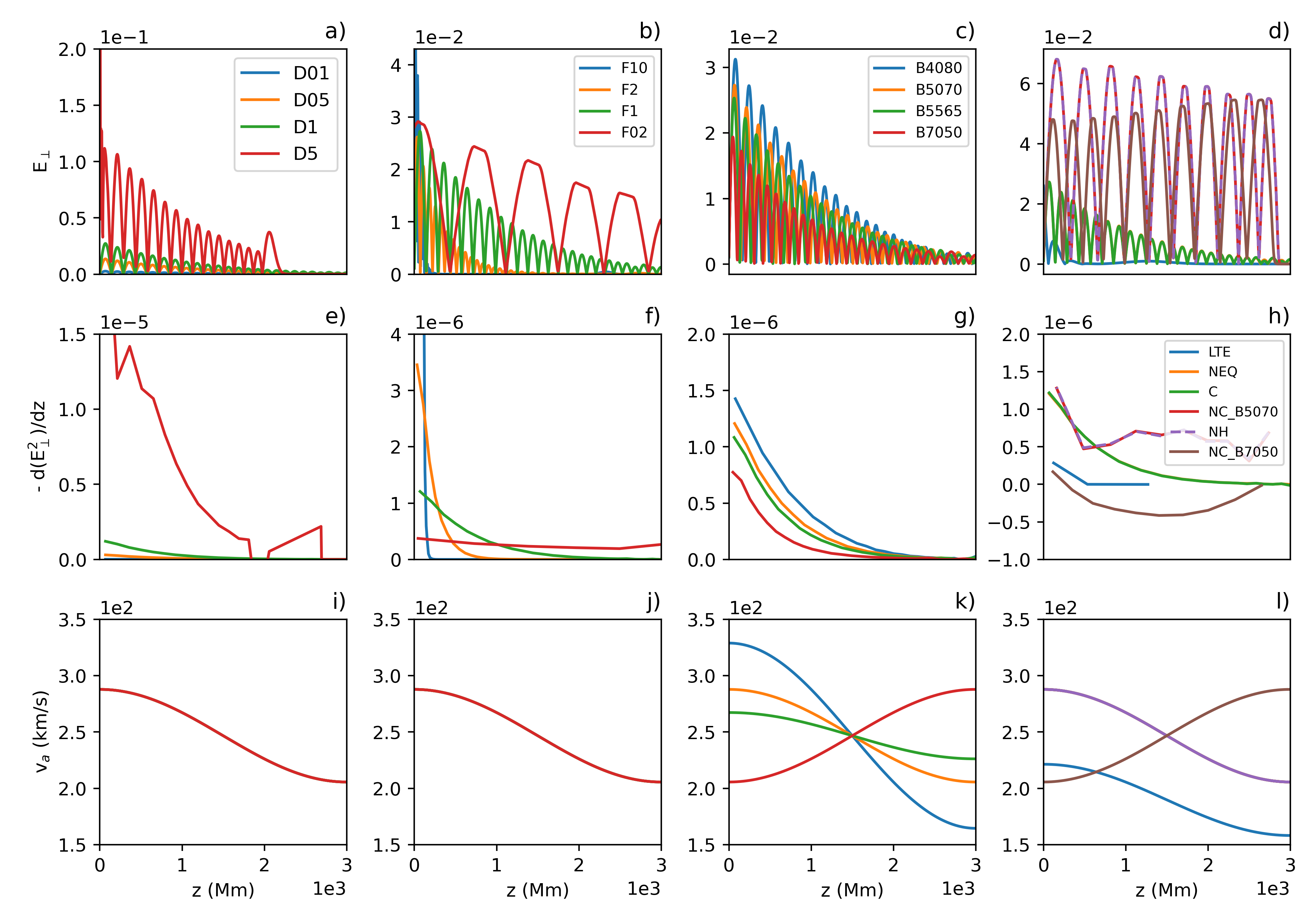}
	\caption{\label{fig:eperp} Electric field perpendicular to the magnetic field (top row), $- dE^2_{\bot}/dz$ (middle row) and Alfvén speed (bottom row) as a function of $z$-axis. We show this for a variety of simulations with varying amplitude, and varying frequency of the Alfvén wave driver, the variation of the guide magnetic field, and different physical assumptions, as shown, respectively, from left to right.}
\end{figure*}

\begin{itemize}
\item The greater the amplitude of the Alfvén wave (panels a and b in Figure~\ref{fig:slopes_t}), the larger the values in the estimated ponderomotive force ($-dE^2_{\bot}/dz$) (panel e of Figure~\ref{fig:eperp}). 
\item The higher the frequency of the Alfvén wave (panels c and d in Figure~\ref{fig:slopes_t}), the larger the values in the estimated ponderomotive force (panel f of Figure~\ref{fig:eperp}). Note that the force drops faster with $z$, when the frequency is higher. This is because of the strong dissipation. 
\item Due to collisions, $E_{\bot}$ decreases with $z$ for any background field configuration considered here ($\Delta B$). This results in a positive ponderomotive force (i.e., in the direction of wave propagation) for constricting and expanding loops. There is a decrease in the acceleration and ponderomotive force when increasing the expansion of the field lines, i.e., with decreasing $\Delta B$. Still, the presence of collisions has a stronger impact on the ponderomotive force and acceleration than the variation of the field configuration and Alfvén speed (panel k). 
\item The collisionless cases reveal the large role of collisions. Indeed, for an expanding or constricting loop, the sign of the acceleration changes, being positive for constricting loops ($\Delta B > 0$), and negative for expanding loops ($\Delta B < 0$) (dots in panels e-h in Figure~\ref{fig:slopes_t}). This is in agreement with the $E_{\bot}$ dependence with $z$, i.e., $E_{\bot}$ decreases in the direction of the propagating wave (with $z$) for a constricting loop (red and purple lines in panels d, h, and l in Figure~\ref{fig:eperp}) (since the Alfvén speed increases) and as a result the estimated ponderomotive force is positive. Whereas for an expanding loop, $E_{\bot}$ increases in the direction of the propagating wave (brown lines panels d, h and l in Figure~\ref{fig:eperp}) (since the Alfvén speed decreases) and, as a result, the estimated ponderomotive force is negative. Note that the sign of this acceleration follows Eq.~\ref{eq:pond1}. %So, constricting loops should lead to an inverse FIP effect for a wave coming from the chromosphere, but this becomes very difficult when collisions are taken into account. 
\item The initial setup of the density population (NEQ vs. LTE) greatly impacts the electric field changes with $z$, hence the estimated ponderomotive force and acceleration of the various fluids, as shown previously. 
\item The motions in the perpendicular plane (and electric field) are very similar between the NC\_B7050\_F1\_D1 case and NH\_B7050\_F1\_D1, i.e., with and without the Hall term. 
\end{itemize}

\begin{figure}
    \includegraphics[width=0.48\textwidth]{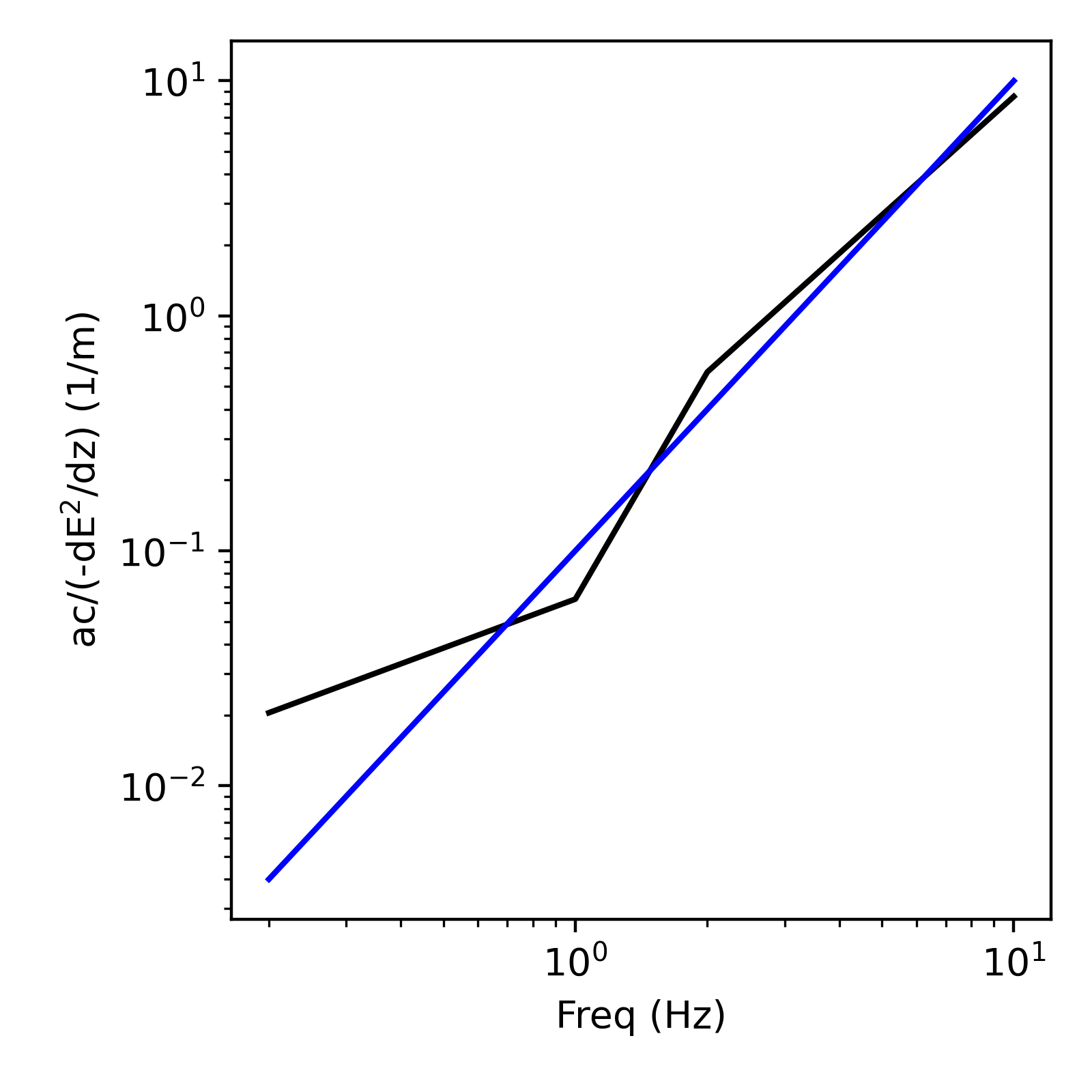}
	\caption{\label{fig:slopes_ratio_ac_de} The ratio of the acceleration with $dE^2_{\bot}/dz$ for ionized helium as a function of frequency. The blue line follows a dependence of $\omega^2$ and matches what is expected from the ponderomotive force associated with damping of Alfv\'en waves from ion-neutral damping (see text for details).}
\end{figure}

It is clear that collisions play a critical role in the ponderomotive force within the parameter range considered here (Tables~\ref{tab:pop} and~\ref{tab:1ds}), which is inspired by chromospheric values. This is shown in the right columns in Figures~\ref{fig:slopes_z}, \ref{fig:slopes_t} and \ref{fig:eperp}). The sign of the ponderomotive acceleration agrees with that found in Eq. \ref{eq:pond1}. However, as shown in Figure~\ref{fig:slopes_ratio_ac_de}, the ratio of the acceleration with (-$dE^2_{\bot}/dz$) does not follow $1/\omega^2$, but follows $\omega^2$. This dependence is, however, in agreement with the predictions from damping of Alfv\'en waves from ion-neutral interactions \citep{Haerendel1992Natur.360..241H,De-Pontieu:1998lr}. 
The predicted damping length $L$ for the wave field is given in Eq.~\ref{eq:damp} and can be compared with the distance over which the wave field $E_{\bot}^2$ decays in the top row of Fig.~\ref{fig:eperp}. The damping length in our simulations is similar to what is predicted and shows only a discrepancy of $\sim 2$. %This could be for a variety of reasons:  
%However, the relation is not matching the factor. We computed the collision frequency of neutral hydrogen with any other ion and, taking into account the magnetic field configuration, and Alfvén speed, found a mismatch of a factor of $\sim3$. This is not surprising since 1) the variation of the wave power in space is not constant or linear, 2) we are not considering assumptions such as XXX 3) we have treated self-consistently a multi-fluid and multi-species environment, and finally, 4) the highest frequency (and highest amplitude) cases reveal some non-neglected non-linearity effects. 

Note that the setup of the simulations allows considering scenarios where the Alfvén wave is driven from ``above", i.e., corona, or ``below", the photosphere. Under those highly simplified scenarios, i.e., no density stratification nor reflections and in the collisionless case, for a wave coming from the corona, an expanding loop with height, i.e., a constricting loop in the direction of the wave propagation, will produce an acceleration opposite to the wave propagation direction, i.e., the FIP effect. In contrast, a constricting loop with height (which may be quite rare, as it is not a condition thought to commonly occur), i.e., an expanding loop in the direction of a wave coming from the corona, will produce an acceleration in the direction of the wave, which could conceivably lead to an inverse FIP effect. However, collisions will lead to a ponderomotive force in the direction of the wave propagation independently of the field configuration, producing accelerations in the direction of the wave propagation. This could imply an inverse FIP effect for a wave coming from the corona. 

\section{Discussions and Conclusions} \label{sec:con}

One of the most generally accepted models to explain the FIP effect is based on the effects of the ponderomotive force, using semi-empirical models \citep[e.g.,][]{Laming:2004qp}. In this study, by combining \iris\ observations with single and multi-fluid models, we go beyond the limitations of those assumptions. We investigate the possible role of the ponderomotive force on accelerating the plasma by considering multi-fluid effects and the ionization under LTE and NEQ conditions. 

Being aware of complexity of the mixture of waves in the solar atmosphere and LOS effects \citep{Okamoto:2011kx,De-Pontieu:2004hq}, we used non-thermal velocities from a transition region line, i.e., \ion{Si}{4} from \iris\ observations to provide constraints on the properties of unresolved high-frequency waves. Assuming that the non-thermal broadening can be ascribed to waves, one can estimate their lowest frequencies and highest amplitudes. We found that plage and network field regions experience the largest non-thermal 
velocities. 

Using the radiative-MHD numerical model described in \citet{Martinez-Sykora:2020ApJ...889...95M}, we investigate the properties of the regions of interest where low and high FIP elements first ionize in the solar atmosphere for a plage region. Our analysis reveals that, in the chromosphere, the largest perturbations of the electric field component perpendicular to the magnetic field occur in areas of enhanced network field or plage regions, associated with type II spicules and large changes in the magnetic field topology. Chromospheric spicules occur most frequently around the network field that is concentrated at the edges of supergranular cells, i.e., where non-thermal broadening in the low TR is enhanced \citep[see also][]{2014ApJ...792L..15P}. In simulations of spicules, they are associated with strong currents and drive Alfvén waves \citep{Martinez-Sykora:2017sci}, suggesting that, if the ponderomotive force indeed drives the FIP effect, this could lead to chemical fractionation. %Therefore, the drivers of type II spicules may be connected to the formation of switchbacks and slow solar wind. 
Our results reveal that the NEQ effects largely impact the region where the chemical fractionation may occur within the chromosphere and needs to be considered when studying the FIP effect. We note that the recent work by \citet{Wargnier2022arxiv.2211.02157} provides another possible chemical fractionation mechanism that depends on the history of the multi-fluid evolution for reconnection events. 

In the last part of this manuscript, inspired by our observational and single fluid analysis, we performed multi-fluid numerical models of Alfvén waves in a non-uniform magnetic field configuration. One has to keep in mind that the multi-fluid models used here are highly simplified: they do not include density stratification or gravity, nor do they capture the chromospheric dynamics, ionization/recombination, thermal conduction, or reflections of waves. Further work is obviously needed to investigate further the various effects that were ignored in this initial experiment. 

However, in our results, the presence of different fluids (ionized and neutral fluids for various species) leads to both collisional and electric coupling. This, in turn, causes a (wave) phase offset between the various fluids and results in damping of the Alfvén waves due to collisions while propagating through the atmosphere. This coupling and damping leads to a positive ponderomotive acceleration, i.e., in the direction of the propagating wave. If the wave propagates upward, the ponderomotive force is upward, and for downward propagating waves, the force is downward. The role of collisions is much stronger than the variations of the magnetic field (and thus the Alfv\'en speed) along the loop, which are meant to mimic a loop that expands or constricts with height. Our results suggest that the presence of ion-neutral damping in the chromosphere has a large potential to dominate any other type of ponderomotive forcing, and may invalidate previous approaches that are based on other forms of the ponderomotive force. For example, the direction of the ponderomotive force in our simulations is different from what has been proposed in the recent literature (see below). Our results suggest that if the ponderomotive force associated with Alfvén waves were the dominant mechanism driving the FIP effect, the Alfv\'en waves should propagate from the chromosphere to corona. This is contrast to what has been assumed so far in current theoretical models of the ponderomotive force \citep{2017ApJ...844..153L}. However, it brings that theory in better alignment with the current observational evidence that indicates predominant upward propagation from the chromosphere into the corona \citep{Okamoto:2011kx} and a dominance of upward propagating waves within the coronal volume at lower frequencies \citep{Tomczyk:2007vn}. 

We find that in our models, the ponderomotive force increases with increasing wave amplitude or frequency of the Alfvén wave. We also find an increase in ponderomotive force in an environment in which the magnetic field strength increases in the direction of the wave propagation. We run cases assuming different approaches to ionizations with both LTE populations and NEQ populations. The two scenarios are very different, leading to different ponderomotive accelerations and damping mechanisms. The LTE case tends to damp the waves faster due to the larger presence of neutrals. 

Our results lead us to speculate about what this means for sunspots, which have much lower densities than other solar regions or phenomena. We speculate that this would lead to a reduced collisional coupling, so that our collisionless case might apply. In that case, this could potentially explain the inverse FIP effect that is sometimes seen in sunspots \citep{Doschek:2016ik}. However, this speculation needs further investigation within the parameter range of a sunspot stratification. 

We finish with the caveat that this first exploration of the relative motions and accelerations of ions with different first ionization potential is highly simplified and does not yet include various effects, including stratification. The latter has the potential to lead to larger spatial variations in the Alfv\'en speed than we considered here, which were only based on the varying guide field (not the density). At a minimum, this would mean that the conditions that we have called ``expanding" or ``constricting" loops (purely based on guide field changes) may well change meaning when stratification is included, given the expected change in sign of $d v_A / d s$. This is important to note since the ponderomotive force also depends on changes in $v_A$. In any case, large changes in Alfv\'en speed have the potential to significantly change the ponderomotive force, including through reflections from the transition region. Future numerical multi-fluid models that include more complexity are needed to move beyond the simplified analytical approaches of the current state-of-the-art FIP models and better understand how the ponderomotive force acts in the low solar atmosphere and can lead to fractionation. 

It is clear from our results however that the presence of collisions in the partially ionized chromosphere introduces a ponderomotive force that, for the upper chromospheric thermodynamic and magnetic field conditions we have considered here, may well dominate other types of ponderomotive force.

\acknowledgements{\longacknowledgment} 

\bibliographystyle{aasjournal}
\bibliography{aamnemonic,collectionbib}

\end{document}